\documentclass[aps,prb,10pt,twocolumn,amsmath,amssymb,longbibliography,superscriptaddress,nofootinbib]{revtex4-2}
\usepackage{amsmath,amssymb,amsfonts} 	% Typical maths resource package
\usepackage{graphicx}
\usepackage{hhline}
\usepackage[latin1]{inputenc}
\usepackage{subfigure}
\usepackage[normalem]{ulem}
\usepackage{xcolor}

\begin{document}
\title{Minority magnons and mode branching in monolayer Fe$_3$GeTe$_2$}

\author{Thorbj\o rn Skovhus}
\email{thosk@dtu.dk}
\affiliation{CAMD, Department of Physics, Technical University of Denmark, 2800 Kgs. Lyngby Denmark}
\author{Thomas Olsen}
\affiliation{CAMD, Department of Physics, Technical University of Denmark, 2800 Kgs. Lyngby Denmark}

\begin{abstract}
We predict the presence of minority magnons in monolayer Fe$_3$GeTe$_2$ using first principles calculations. Minority magnons constitute a new type of collective magnetic excitation which increase the magnetic moment---contrary to ordinary (majority) magnons which lower it---giving rise to spin-raising poles in the dynamic susceptibility $\chi^{-+}(\omega)$.
The presence of such quasi-particles is made possible by the nontrivial ferromagnetic band structure of Fe$_3$GeTe$_2$. %originating from its nonequivalent Fe sublattices. %The signature of minority magnons is 
%The result is a strong peak in the dynamic spin-raising susceptibility $\chi^{-+}(\omega)$ in the long wavelength limit, which is the hallmark of minority magnon physics.
We calculate the susceptibility using time-dependent density functional theory and perform a detailed mode analysis, which allows us to identify and investigate individual magnon modes as well as the Stoner excitations that constitute the many-body spectrum. %For minority as well as majority magnons, 
The analysis reveals a plethora of both majority and minority excitations, which in addition to the main magnon branches include both satellite, valley and spin-inversion magnons thanks to the electron itinerancy of the system. 
Crucially, the analysis allows us to separate peaks of a coherent collective nature from those of mixed magnon/Stoner nature. To this end, we predict that the lowest energy minority magnon mode of monolayer Fe$_3$GeTe$_2$ indeed constitutes a coherent collective quasi-particle at long wavelengths and introduce a simplistic gaussian model for the observed minority magnon enhancement. 
The underlying physics is in no way restricted to Fe$_3$GeTe$_2$, and minority magnons are thus expected to be observable in other complex ferromagnetic materials as well.
%Sed eu venenatis ex, sit amet malesuada lorem. Pellentesque consequat nisl at risus iaculis auctor. Aliquam purus magna, sagittis sit amet enim ac, venenatis mollis urna. Nulla maximus sagittis risus, a porta velit feugiat a. Aliquam justo ante, pellentesque eget malesuada eu, eleifend finibus lacus. Aenean placerat magna ut molestie aliquam. Duis suscipit dictum luctus. Nulla quis ligula felis. Aliquam pharetra metus est, sit amet tempor mauris tempus sit amet. Fusce posuere nisl eu sapien eleifend consequat. Pellentesque. 
\end{abstract}
\maketitle

%%%%% Background %%%%
% Here we introduce the magnon quasi-particles and present the historical spin-wave picture in the context of the Heisenberg model. We explain why such a description does not host excited states with raised spin angular momentum in ferromagnets and the reasons to doubt its validity for itinerant systems.

%Magnetic crystals are characterized by exhibiting finite magnetization at zero temperature in the absence of external magnetic fields. The finite magnetization of the ground state entails the existence of low energy magnetic excitations, which for systems with weak spin-orbit coupling alter the local direction of magnetization, without changing the charge distribution nor the size of the magnetization (to lowest order). In a quasi-particle picture, these excitations are referred to as magnons, carrying well defined crystal and spin angular momenta. 

\section{Introduction}

In many aspects, the dynamics of magnetic crystals is governed by the excited states which alter the local direction of magnetization. For example, the critical temperature at which magnetic order is lost can largely be understood as the point where the magnetization vanishes due to their thermal population. These so-called transverse magnetic excitations can roughly be divided into two types, Stoner pairs and magnons, 
%in magnetic crystals are referred to as magnons, 
each carrying exactly one unit of spin angular momentum in the absence of spin-orbit coupling.
Stoner pairs are electron-hole pairs of a single-particle nature, forming a continuum out of the band transitions where an electron from an occupied band is excited above the Fermi level to a band of the opposite spin. Due to the exchange splitting in magnetic systems, the Stoner continuum is gapped at zero momentum transfer, remaining so throughout the Brillouin Zone (BZ) for magnetic insulators and half-metals.
Magnons, on the other hand, are of a collective nature, that is, the alteration of the magnetization direction is mediated by a collection of correlated electrons, rather than of a single (dressed) electron-hole pair. The correlation is dominated by exchange and thus attractive, permitting magnon excitations to exist in the energy gap below the Stoner continuum, with an acoustic magnon of vanishing energy in the $q\rightarrow 0$ limit guaranteed by the Goldstone theorem. 
The correlated nature of magnons makes for a challenging theoretical treatment, and by far the majority of theoretical \cite{Yosida1996,Szilva2023} as well as experimental \cite{Squires1978, Jensen1991} works rely on Heisenberg Hamiltonians to model and interpret magnon excitations. In this picture, the magnetic structure is modelled in terms of localized magnetic moments coupled by exchange, and magnons manifest themselves as spin waves. The approach works very well \cite{Boothroyd2020} to describe spin-lowering magnons outside of the Stoner continuum, but does not explicitly account for the Stoner pair excitations and their coupling to the magnons. Furthermore, it bears no account of magnetic excitations that increase the local magnetic moment, since
%ith a single localized moment per magnetic site, 
the spin-raising operator $\hat{S}_i^+$ cannot be applied to sites $i$ where the local spin is aligned with the $z$-axis. %Hence, the quantum mechanical SWEOM for a ferromagnet only includes modes which reduce $S^z$. However, when solved semi-classically, an additional set of eigenmodes that increase $S^z$ appear at negative frequencies, constructed as linear combinations of $S_i^+$.

%%%%% Motivation %%%%%
% Here we present the alternative view of the Stoner model. We introduce the concept of collective enhancement and the influence of minority Stoner pairs with appropriate band structures / density of states.

In metallic magnets, the Stoner continuum becomes gapless for all finite wave vectors $\mathbf{q}$ which connect the majority and minority spin Fermi surfaces. As a result, the magnons inevitably couple directly to the Stoner continuum, in which case the distinction between the two types of excitations becomes less clear. The coupling means that the magnon resonances will be damped (Landau damping). If the Stoner scattering is weak at magnon resonance, the quasi-particle picture remains valid and the magnon simply attains a finite lifetime, but with increasing Stoner scattering intensity, the magnon will gradually be washed out in the spectrum, until no clear magnon peak remains. 
%In this letter, we quantify this continuous transition between the limiting magnon/Stoner classifications via the collective self-enhancement function for each magnon mode, allowing us to separate magnon features which are coherent, but with a finite lifetime, from those which are spectrally indistinguishable from the former, but not fully self-sustained in the electronic correlations due to the admixture of Stoner pair character.
Furthermore, small spectral features in the Stoner continuum can have a massive influence on the magnon lineshape. If a magnon resonance and a Stoner feature cross, it leads to a discontinuity in the magnon dispersion, with both magnon branches coexisting for wave vectors $\mathbf{q}$ in the vicinity of the crossing. Despite being absent from the Heisenberg model, such effects are consistently predicted at higher levels of theory for simple ferromagnets such as Fe and Ni \cite{Savrasov1998,Karlsson2000,Buczek2011b,Rousseau2012,Cao2017,Singh2018,Okumura2019,Friedrich2020,Skovhus2021}, where Ni is particularly well-known for its pronounced magnon branching, also experimentally \cite{Mook1985}. The potential richness in complex multisublattice itinerant ferromagnets is even more astounding. Particularly, we will entertain the possibility of forming spin-raising magnons out of the spin-raising part of the Stoner continuum, involving transitions from occupied minority $d$-states to unoccupied majority $d$-states. We will refer to these as minority magnons. 
%For metals, it is not obvious that the picture of localized moments remains valid, and indeed there are several well known itinerant electron effects, such as Stoner stripes and Landau damping, which cannot be described in the Heisenberg model. Instead one needs an explicit treatment of the electronic bands, such as in the Stoner model. Here, a Hubbard model is treated in the mean field, resulting in a spin splitted band structure for certain ranges of parameters. Due to the spin splitting there exists a continuum of electron-hole pairs with opposite spin (Stoner pairs), which is gapped at zero momentum transfer. The magnons are modelled as coherent superpositions of the majority part of the continuum, that is, of the pairs where an electron has been transferred from an occupied majority spin state to an unoccupied minority spin state, solving the Stoner equation of motion in the random phase approximation (RPA). In fact, the dispersion of the collective magnon mode can be extracted from the RPA susceptibility as poles on the real frequency axis situated in the gap below the Stoner continuum \cite{Moriya1985,Yosida1996}. In addition to the majority Stoner pairs, which lower $S^z$, the Stoner continuum can also include minority Stoner pairs involving an occupied minority spin state and an unoccupied majority spin state, thus raising $S^z$. 
The possible presence of such excitations is usually neglected, and for good reason. 
In single-band models of itinerant ferromagnetism like the spin-polarized homogeneous electron gas \cite{Moriya1985,Yosida1996}, minority Stoner pairs only occur once the Stoner gap closes. %for wave vectors large enough to connect the majority and minority Fermi surfaces. 
As a result, there is no minority Stoner continuum at vanishing $q$ and coherent minority magnons do not form \cite{Friedrich2020}. However, in a multi-band ferromagnet, there is nothing prohibiting the existence of such a minority magnon mode. The only requirement is that a ferromagnet has at least one occupied minority $d$-band as well as available majority $d$-states above the Fermi level, giving rise to a gapped minority Stoner continuum at $q=0$.

%%%%% Introduction %%%%%
% Here we present our prediction of the existence of minority magnons based on a first principles description. Furthermore, we tease that we have found such minority magnon modes in FGT along with pronounced itinerant electronic effects in the form of mode branching of the optical majority modes and provide an extremely brief background on previous FGT research.

In this paper, we showcase how minority magnons enter the first principles theory of linear response and investigate their presence in the itinerant ferromagnetic monolayer Fe$_3$GeTe$_2$ \cite{Fei2018}. This compound has recently spurred significant interest due to its gate-tunable Curie temperature (exceeding room temperature) as well as efficient spin-orbit torque switching \cite{Alghamdi2019, Wang2023} making it a highly promising venue for 2D spintronics applications. The presence of three Fe atoms (two of which are equivalent) in the unit cell implies a rich electronic structure and, as we will see, gives rise to minority magnon excitations.  We base our analysis on calculations of the Fe$_3$GeTe$_2$ susceptibility within the adiabatic local density approximation (ALDA) to linear response time-dependent density functional theory (LR-TDDFT) \cite{Runge1984,Gross1985,Savrasov1998,Buczek2011b,Rousseau2012,Singh2018,Skovhus2021} and present a simplistic model for the minority magnon enhancement in Fe$_3$GeTe$_2$, illustrating how collective minority magnons form, in Fe$_3$GeTe$_2$ and in other complex ferromagnets. In addition, we also provide a detailed account of the pronounced mode branching of the optical majority modes in Fe$_3$GeTe$_2$ and discuss how one can distinguish magnons from Stoner pairs in practice and quantify the coherency of collective excitations.

%%%%% Theory %%%%%
% Here we present the ab initio theory, which allows for the existence of minority magnons with the proper density of states. We introduce the self-enhancement function and the Dyson equation and justify using the Xi for characterization of the resonances.

\section{Linear response theory}

For ferromagnets with vanishing spin-orbit coupling, a perturbing transverse magnetic field induces a transverse magnetization without coupling to the longitudinal degrees of freedom to linear order \cite{Buczek2011b,Skovhus2021}. With the ground state spin-polarized along the positive $z$-direction,
\begin{equation}
    \delta n^\alpha(\mathbf{r}, t) = \sum_\beta \int_{-\infty}^\infty dt' \int d\mathbf{r} \: \chi^{\alpha\beta}(\mathbf{r},\mathbf{r}',t-t') W_\mathrm{ext}^\beta(\mathbf{r}', t'),
\end{equation}
where $n^\alpha$ is the transverse electron spin-density, $\alpha,\beta\in\{x,y\}$, and $\mathbf{W}_\mathrm{ext}=\mu_\mathrm{B} \mathbf{B}_\mathrm{ext}$. Crucially, the four entries of the transverse magnetic susceptibility can be decomposed into circular coordinates,
\begin{equation}
    \chi^{[x,y]} = 
    \begin{pmatrix}
        \chi^{+-} + \chi^{-+} & i\chi^{+-} - i\chi^{-+} \\
        -i\chi^{+-} + i\chi^{-+} & \chi^{+-} + \chi^{-+}
    \end{pmatrix},
\end{equation}
where $\chi^{+-}(\omega)$ and $\chi^{-+}(\omega)$ yield the response to negatively and positively circulating magnetic fields, $\mathbf{B}_\mathrm{ext}(\mathbf{r},t) = B_\mathrm{ext}(\mathbf{r}) [\cos(\omega t) \mathbf{e}_x \mp \sin(\omega t) \mathbf{e}_y]$. The dissipative part of the response (the imaginary part of the frequency dependence) yields the so-called scattering function $S^{+-}(\omega)$, characterizing the inherent inelastic scattering amplitude of the two circularities at positive and negative frequencies respectively 
\cite{VanHove1954,Kubo1966,Jensen1991}.
%
% \begin{equation}
%     S^{+-}(\mathbf{r}, \mathbf{r}', \omega) = - \frac{1}{2\pi i} \left[\chi^{+-}(\mathbf{r}, \mathbf{r}', \omega) - \chi^{-+}(\mathbf{r}', \mathbf{r}, -\omega)\right],
%     \label{eq:S+- dissipative part}
% \end{equation}
%
%The scattering function %is defined in terms of the dissipative part of the circular susceptibilities and 
Essential to this work, $S^{+-}$ can be decomposed into two distinct spectral functions \cite{Skovhus2021}: 
\begin{equation}
    S^{+-}(\mathbf{r}, \mathbf{r}', \omega) = A^{+-}(\mathbf{r}, \mathbf{r}', \omega) - A^{-+}(\mathbf{r}', \mathbf{r}, -\omega).
    \label{eq:S+- spectral functions}
\end{equation}
These are spectral functions for the excited states $|i\rangle$ which decrease and increase $S^z$ by $\hbar$ respectively, weighted by the matrix elements $n^\pm_{ii'}(\mathbf{r}) \equiv \langle i| \hat{n}^\pm(\mathbf{r}) |i'\rangle$:
\begin{equation}
    A^{+-}(\mathbf{r}, \mathbf{r}', \omega) = \sum_{i>0} n^+_{0i}(\mathbf{r}) n^-_{i0}(\mathbf{r}')\, \delta\hspace{-1pt}(\hbar\omega - (E_i - E_0)).
    \label{eq:A+- definition}
\end{equation}
Here $\hat{n}^+(\mathbf{r})=\hat{\psi}^\dagger_\uparrow(\mathbf{r})\hat{\psi}_\downarrow(\mathbf{r})$, $\hat{n}^-(\mathbf{r})=\hat{\psi}^\dagger_\downarrow(\mathbf{r})\hat{\psi}_\uparrow(\mathbf{r})$ and $\pm$ indices can be interchanged to yield $A^{-+}$. Thus, by definition any (bright) majority excitation will manifest itself as positive spectral weight in the scattering function $S^{+-}$ at nonnegative frequencies, whereas minority excitations provide negative spectral weight at negative frequencies.

\subsection{LR-TDDFT}

Within the framework of LR-TDDFT, the many-body susceptibility $\chi^{+-}$ can be calculated from first principles based on the corresponding $\chi^{+-}_\mathrm{KS}$ of the noninteracting Kohn-Sham system by inverting a Dyson-like equation \cite{Gross1985}. For the work presented here, we have used a new implementation \cite{new_implementation} of the ALDA in the GPAW linear response code \cite{Skovhus2021,Yan2011,mortensen2024}, which enables us to accurately describe magnons in 2D materials. Here, the Fourier transform of the ALDA self-enhancement function,
\begin{equation}
    \Xi_\mathrm{ALDA}^{++}(\mathbf{r},\mathbf{r}',\omega) = \chi_\mathrm{KS}^{+-}(\mathbf{r},\mathbf{r}',\omega) f_\mathrm{LDA}^{-+}(\mathbf{r}'),
    \label{eq:ALDA Xi}
\end{equation}
is calculated using the projector augmented wave method \cite{Blochl1994,Skovhus2021}, and the Dyson equation is inverted in the plane-wave basis:
\begin{equation}
    \chi^{+-}(\mathbf{q},\omega) = \left[1 - \Xi^{++}(\mathbf{q},\omega)\right]^{-1} \chi_\mathrm{KS}^{+-}(\mathbf{q},\omega).
    \label{eq:Dyson}
\end{equation}
In previous implementations \cite{Skovhus2021}, $\chi_\mathrm{KS}^{+-}$ and $f_\mathrm{LDA}^{-+}$ were Fourier transformed individually, which is highly problematic for 2D materials, since the LDA kernel diverges in vacuum. In contrast, the self-enhancement function \eqref{eq:ALDA Xi} is exponentially localized to the monolayer, making its Fourier transform well behaved. %Noting that $f_\mathrm{LDA}^{-+}(\mathbf{r})=f_\mathrm{LDA}^{+-}(\mathbf{r})$, $\pm$ indices can be interchanged in Eqs. \eqref{eq:ALDA Xi} and \eqref{eq:Dyson} to obtain the Dyson equation for $\chi^{-+}$.

%%%%% Results %%%%%
% Here we present and discuss the results and note the computational details as we go.

\section{Results and discussion}

\subsection{Ground state and band structure}

\begin{figure*}[t]
    \centering
    \includegraphics[scale=1.0]{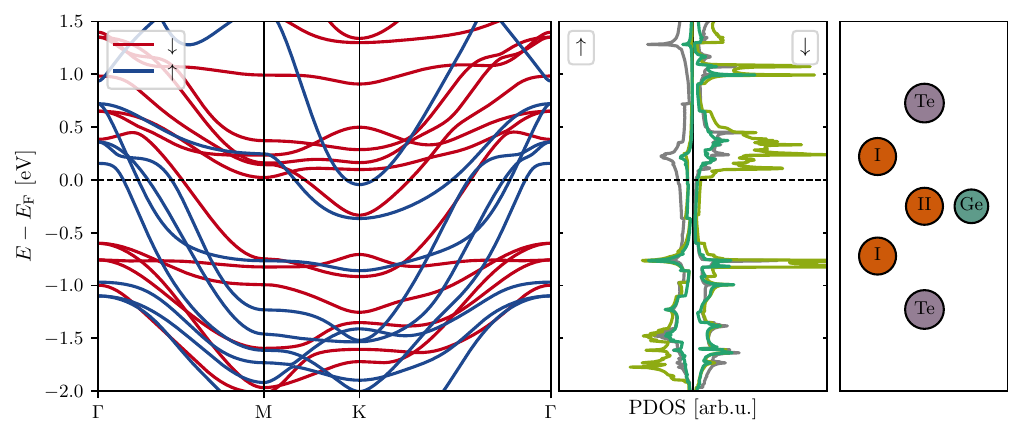}
    \caption{LDA band structure and projected density of states of the Fe$_3$GeTe$_2$ monolayer, along with a side view of the atomic arrangement inside the unit cell (for a top view of the crystal structure, see Fig. \ref{fig:Fe3GeTe2_crystal_structure} in Appendix \ref{sec:computational details}). Projections onto Fe(I) $d$-orbitals are depicted in green, Fe(II) $d$-orbitals in teal and the remaining density of states in grey.}
    \label{fig:Fe3GeTe2_crystalstructure_bandstructure_and_pdos}
\end{figure*}
Calculation of the ALDA susceptibility takes the LDA ground state as a starting point. The unit cell of monolayer Fe$_3$GeTe$_2$ contains three Fe atoms, see Fig. \ref{fig:Fe3GeTe2_crystalstructure_bandstructure_and_pdos}, out of which we find the two equivalent Fe(I) atoms to carry a magnetic moment of 2.39 $\mu_\mathrm{B}$ each, whereas the third Fe(II) atom has a reduced moment of 1.36 $\mu_\mathrm{B}$. This is easily rationalized based on the LDA band structure and projected density of states presented in Fig. \ref{fig:Fe3GeTe2_crystalstructure_bandstructure_and_pdos}. 
Here the occupied density of minority $d$-states is dominated by several weakly dispersive bands with roughly equal weight on the $d$-orbitals of the Fe(I) and Fe(II) atoms. Since there are two of the former, this quenches the magnetic moment of the Fe(II) atom in comparison. In addition to the reduced moment, we identify in the band structure also an almost flat minority band 0.71--0.83 eV below the Fermi level giving rise to a narrow and high density of minority $d$-states. Along with the majority spin bands of partial $d$-orbital character that penetrate the Fermi level, this creates the basis for a gapped minority Stoner continuum in the long wavelength limit, thus making monolayer Fe$_3$GeTe$_2$ a promising candidate material to host minority magnons. 
%
%As described in the main text, the low energy minority magnon mode arises due to a peak at the onset of the minority Stoner continuum $S^{+-}_\mathrm{KS}(\mathbf{q},\omega)$, situated around $-1.2$ eV for $\mathbf{q}=\mathbf{0}$. 
%
%Based on the projected density of states, it seems that the main hole suppliers to these specific minority Stoner pairs are the flat minority band 0.71--0.83 eV below the Fermi level and the minority band which lies (rather nondispersively) just below it at the M-point, 0.99 eV below the Fermi level. 
%The majority electrons of the Stoner pairs are supplied by the small selection of bands with $d$-orbital weight that penetrate the Fermi level, primarily providing their projected density of $d$-states around either the M-point or close to the center of the BZ.

\subsection{Transverse magnetic excitations in the long wavelength limit}

\begin{figure}[tb]
    \centering
    \includegraphics[scale=1.0]{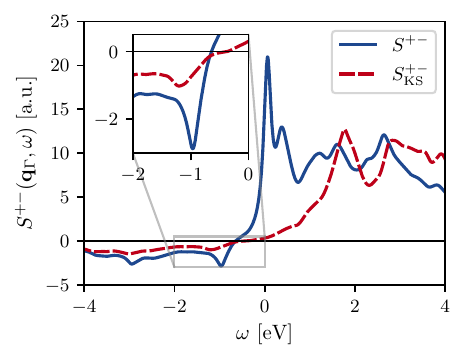}
    \caption{Transverse magnetic scattering function for the many-body and Kohn-Sham systems in the long wavelength limit, with majority and minority excitations manifesting themselves at positive and negative frequencies respectively.}
    \label{fig:Fe3GeTe2_total_scattering}
\end{figure}
In Fig. \ref{fig:Fe3GeTe2_total_scattering}, we show the full trace of ALDA scattering function $S^{+-}$ evaluated at %$\mathbf{q}=\mathbf{0}$
$\mathbf{q}_\Gamma$ and compare it to the %trace of the corresponding scattering function calculated based on $\chi^{+-}_\mathrm{KS}$. 
scattering function of the Kohn-Sham system. At negative frequencies, we find the minority excitations. 
%As a direct consequence of the intricate Fe$_3$GeTe$_2$ band structure, 
$S^{+-}_\mathrm{KS}$ displays a wide continuum of noninteracting minority Stoner pairs, which is gapped and exhibits a somewhat broad peak at its onset. %, due in part to the nondispersive minority band below the Fermi level \cite{Supplementary}. 
%For the many-body scattering function, the electron correlations result in at least two distinct peaks inside the minority Stoner continuum, both representing collective minority magnon modes. Interestingly, the correlation effects increase the (absolute) spectral weight at negative frequencies. 
\begin{figure*}[tb]
    \centering
    \includegraphics[scale=1.0]{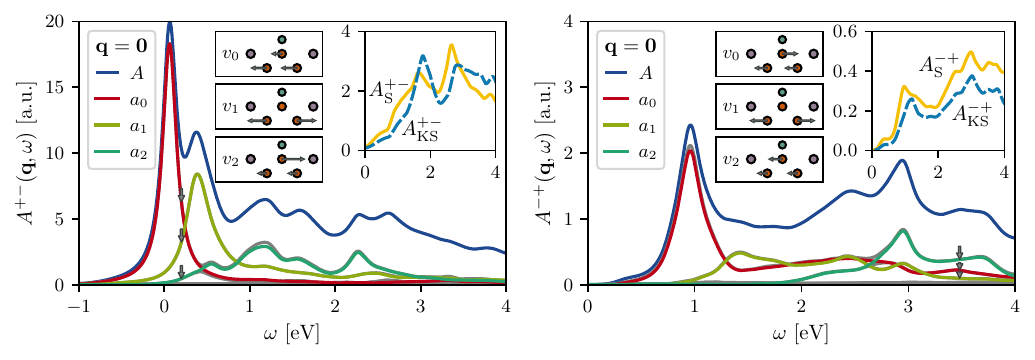}
    \caption{Eigenvalue decomposition of the spectrum of majority (left) and minority (right) excitations in the long wavelength limit. The spectral function $A(\mathbf{q},\omega)$ (blue) %is constructed as the sum of positive eigenvalues (grey) of the corresponding scattering function $S(\mathbf{q},\omega)$. The 
    is given by the sum of all eigenvalues (grey), and the
    magnon lineshapes $a_n(\mathbf{q},\omega)$ (lying essentially on top of the three largest eigenvalues) are %extracted as expectation values of the spectral function using the 
    calculated as spectral projections onto eigenvectors extracted at the shown frequency (vertical arrows). %The atomic projections of each magnon eigenmode are shown in the insets at the center. 
    The magnon eigenmodes are shown in the central insets (Fe in orange, Ge in teal and Te in purple).
    %The sum of remaining positive eigenvalues is identified as the many-body Stoner spectrum and is compared to the sum of positive eigenvalues for the Kohn-Sham scattering function $S_\mathrm{KS}(\mathbf{q},\omega)$ in the upper right inset.
    In the upper right insets, the remaining many-body Stoner spectrum is compared to the spectral function of the Kohn-Sham system, see also Appendix \ref{sec:computational details} for computational details.}
    \label{fig:Fe3GeTe2_mode_extraction}
\end{figure*}
In comparison, a much clearer excitational peak is featured in the many-body scattering function at frequencies around the lower end of the minority Stoner continuum.
%In the many-body scattering function, electron correlations have resulted in a clear collective minority magnon peak below the main Stoner continuum. % (in absolute frequencies). 
In fact, the correlation effects have enhanced the spectral weight across all the calculated negative frequencies.
%Since the spectral weight integrated over all frequencies remain equal 
Since the zeroth moment of the scattering function is identical
in the Kohn-Sham and many-body systems thanks to a sum rule \cite{Rousseau2012,Skovhus2021}, this means that the total spectral weight at positive frequencies must have been increased as well. As we shall see, this is a consequence of the mutual collective enhancement of the majority and minority channels. At the majority side of the Kohn-Sham scattering function, the Stoner continuum features a broad but distinct peak around 1.75 eV corresponding to the exchange splitting of the $d$-bands, but also additional spectral features of a comparable intensity. %not only a single peak (corresponding to a unique exchange splitting of the $d$-bands), but two with a valley inbetween, as well as several other smaller features.
%Also the optical magnon modes extend deeply into the eV frequency range, exhibiting significant mode branching due to complex frequency-dependence of the Stoner continuum. As a result, the many-body scattering function in Fig. \ref{fig:Fe3GeTe2_total_scattering} exhibits many more than the three peaks at positive frequencies expected based on a Heisenberg model treatment (one for each Fe-atom).
The intricate frequency-dependence of the Stoner continuum results in significant branching of the optical magnon modes. For the many-body scattering function in Fig. \ref{fig:Fe3GeTe2_total_scattering}, this means that magnon excitations contribute with more than just the three peaks at positive frequencies expected based on a Heisenberg model treatment (one for each Fe-atom).

%
%To further dissect the scattering function and to analyze each magnon mode independently, we reduce $S^{+-}$ to a plane-wave cutoff of 200 eV and diagonalize it, such that the majority and minority spectral functions \eqref{eq:A+- definition} can be separated in accordance with Eq. \eqref{eq:S+- spectral functions}, that is, in terms of the positive and negative eigenvalues respectively. For each calculated frequency $\omega$ and wave vector $\mathbf{q}$, we save the 27 and 15 largest eigenvalues of $A^{+-}$ and $A^{-+}$ respectively, along with the corresponding eigenvectors and the full trace of eigenvalues. 
%To further dissect the scattering function and 
In addition to magnon excitations, the many-body scattering function presented in Fig. \ref{fig:Fe3GeTe2_total_scattering} also includes the Stoner pairs. To separate collective and single-particle-like excitations, we isolate and plot the majority and minority spectral functions \eqref{eq:A+- definition} according to Eq. \eqref{eq:S+- spectral functions}, that is, in terms of the positive and negative eigenvalues of $S^{+-}$ respectively, see Fig. \ref{fig:Fe3GeTe2_mode_extraction}.
%This allows us %to analyze how each eigenmode contributes to the total spectral function $A(\mathbf{q},\omega)$ and to extract the magnon 
%to identify the lineshapes of the collective modes as shown in Fig. \ref{fig:Fe3GeTe2_mode_extraction}. % for $\mathbf{q}=\mathbf{0}$. 
In contrast to the Kohn-Sham spectrum, the many-body spectral functions exhibit exactly three eigenmodes with scattering amplitudes order(s) of magnitude larger than the rest (which are barely visible in Fig. \ref{fig:Fe3GeTe2_mode_extraction}). 
These are the collective eigenmodes of the system, corresponding directly to the three spin-wave modes of the Heisenberg model, but in this case with complex lineshapes with several spectral features. %driven by the electronic correlations of the system. 
Apart from a difference in scattering amplitude, the picture is strikingly similar for the majority and minority channels, with the notable exception that only the majority channel exhibits a Goldstone magnon resonance (peak at $\omega=0$). Instead, the minority mode with lowest energy is peaked at 0.96 eV in the long wavelength limit. 
% %In order to extract the magnon lineshape of each collective mode, we reconstruct the majority and minority spectral functions \eqref{eq:A+- definition} from the saved eigendecomposition and evaluate the lineshapes 
% We extract the majority magnon lineshapes as the inner product $a_n(\mathbf{q},\omega) = \langle v_n| A(\mathbf{q},\omega) |v_n\rangle$, using magnon eigenvectors $|v_n\rangle$ taken %from a single carefully chosen frequency. Since the magnon eigenvectors change slightly as a function of frequency, this is a somewhat tricky operation. For frequencies where two eigenvalues are close, the eigenvectors can mix leading to features reminiscent of avoided crossings. To extract the majority lineshapes in a reliable fashion we therefore maximize 
% at the frequency $\omega$ which maximizes
% the minimum eigenvalue difference between the three modes: $\max_\omega \left( \min_{n>0} \left[\lambda_{n-1}(\omega) - \lambda_n(\omega)\right]\right)$. 
% This ensures the extraction of well-defined lineshapes, since the eigenvectors are only sensitive to the frequency in the vicinity of crossing eigenvalues.
% For the minority magnons, we choose in this letter to focus on the lowest energy mode, extracting its mode vector at the eigenvalue maximum. The only exception is Fig. \ref{fig:Fe3GeTe2_mode_extraction}, where we showcase that also the three minority magnon mode lineshapes can be extracted in a similar fashion to the majority modes, albeit a little less convincingly so since the $n=2$ mode does not extend to lower frequencies than 2 eV. 
Subtracting the lineshapes of the three collective modes from the trace of the spectral function $A(\omega)$, one is left with the many-body Stoner spectrum $A_\mathrm{S}(\omega)$. In Fig. \ref{fig:Fe3GeTe2_mode_extraction}, we compare this to the Kohn-Sham spectrum, which describes the Stoner pairs at the noninteracting level.
%, calculated as the sum of positive/negative eigenvalues of the 500 eV plane-wave cutoff scattering function. 
The two spectra display remarkably identical features, meaning that the electronic correlations at the ALDA level only lead to a slight renormalization of the Stoner pairs.

\subsection{Collective enhancement and magnon coherency}

\begin{figure*}[tb]
    \centering
    \includegraphics[scale=1.0]{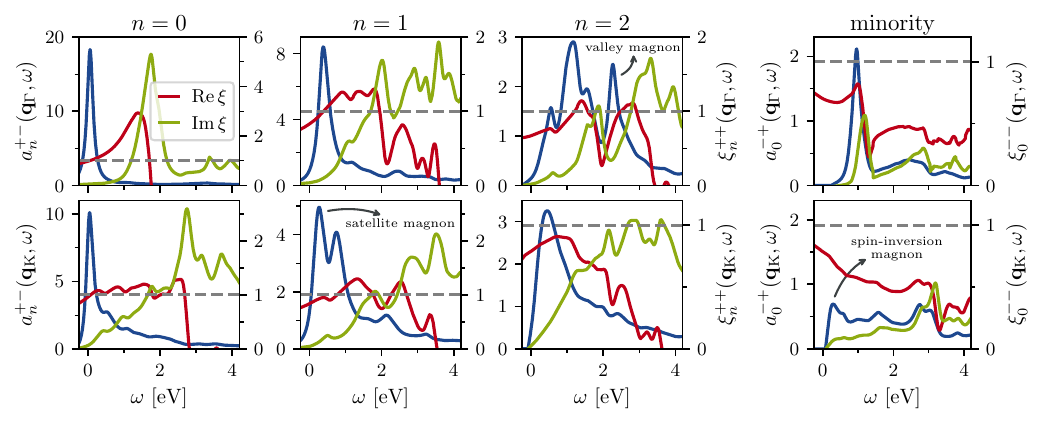}
    \caption{Collective enhancement of the three majority and low energy minority magnon modes %in Fe$_3$GeTe$_2$ 
    (left to right) in the long wavelength limit (top row) and at the $K$ high-symmetry point (bottom row). The magnon lineshapes $a_n(\mathbf{q},\omega)$ (left axis in blue) are the result of an enhancement factor of $\left[1 - \mathrm{Re}\: \xi_n(\mathbf{q},\omega)\right]^{-1}$, where $\xi_n$ (right axis in red and green for the real and imaginary parts) is the projection of the magnon mode onto the corresponding self-enhancement function.}
    \label{fig:Fe3GeTe2_enhancement_analysis}
\end{figure*}
In order to understand which of the peaks in the collective eigenmodes that correspond to coherent magnon quasi-particles, 
%and which that arise due to the direct coupling to a specific Stoner feature, 
we return to the Dyson equation \eqref{eq:Dyson} and the self-enhancement function \eqref{eq:ALDA Xi}. 
%Next, we will analyze the magnon mode enhancement from the Dyson equation \eqref{eq:Dyson} and categorize the different peaks observed within a single mode. 
%Outside the Stoner continuum, the many-body susceptibility $\chi^{+-}$ is pole-free, except for singular points where $\left[1 - \Xi^{++}(\mathbf{q},\omega)\right]\rightarrow 0$. 
$\Xi^{++}(\mathbf{r}, \mathbf{r}', t-t')$ encodes the induced transverse magnetization at time $t$ and position $\mathbf{r}$ due to the fluctuations in the transverse exchange-correlation field that emerge in response to a change in the magnetization at time $t'$ and position $\mathbf{r}'$. If the self-enhancement function attains an eigenvalue of unity, it thus entails the existence of a fluctuational eigenmode $\delta n^\alpha(\mathbf{r},\omega)$ which is self-sustained via the effective electron-electron interaction and gives rise to a pole in the many-body susceptibility as $\left[1 - \Xi^{++}(\mathbf{q},\omega)\right]\rightarrow 0$ in the Dyson equation \eqref{eq:Dyson}. This is what is understood as an undamped magnon resonance in the first principles theory. Inside the Stoner continuum, however, the eigenvalues of $\Xi^{++}$ also attain a finite imaginary part. In fact, the poles of $\Xi^{++}$ and $\chi_\mathrm{KS}^{+-}$ are identical in the ALDA. Nevertheless, %for each pair of unity-crossings (as a function of frequency) that the real part of a $\Xi^{++}$ eigenvalue has ($\Xi$ vanishes away from the Stoner continuum, so crossings always come in pairs), the lower crossing still gives 
the real part of a given $\Xi^{++}$ eigenvalue might still reach unity at a set of specific frequencies giving rise to correlation-driven resonances. The spectral peaks of finite width that follow can still be interpreted as coherent magnon quasi-particles, but in this case with a finite lifetime. If, on the other hand, the real part of a $\Xi^{++}$ eigenvalue approaches unity but reaches a maximum before crossing it ($\Xi^{++}$ vanishes away from the Stoner continuum), the resulting peak in the many-body spectrum cannot be regarded as a coherent collective quasi-particle in the sense that it is not fully self-sustained in the electronic correlations. At the same time, the peak is neither a Stoner pair excitation, since it is absent from the single-particle spectrum and emerges from collective enhancement for the exact same reasons that coherent magnons do. Therefore, we will refer to such peaks as incoherent magnon peaks, well-knowing that they exist as mixtures of collective and single-particle character.
%Nevertheless, if the real part of a $\Xi^{++}$ eigenvalue reaches unity it will still give rise to a magnon resonance and we will interpret the corresponding peak as a coherent magnon quasi-particle of a finite lifetime.

%
In Fig. \ref{fig:Fe3GeTe2_enhancement_analysis}, we showcase these different magnon enhancement scenarios for monolayer Fe$_3$GeTe$_2$ by projecting the self-enhancement function onto the extracted mode vectors, $\xi_n(\mathbf{q},\omega) = \langle v_n| \Xi(\mathbf{q},\omega) |v_n\rangle$.
%, reconstructing $\Xi$ post-calculation from its 42 largest eigenvalues. 
For the Goldstone ($n=0$) mode at %$\mathbf{q}=\mathbf{0}$, 
$\mathbf{q}_\Gamma$,
the majority Stoner spectrum consists of a single main peak in $\mathrm{Im}\:\xi^{++}_0(\omega)$ (from the exchange splitting of the $d$-bands) giving rise to an undamped coherent magnon resonance below it at the point where $\mathrm{Re}\:\xi^{++}_0(\omega)$ crosses unity. At shorter wave lengths, the Stoner continuum extends to progressively lower frequencies and at the K-point the lineshape has undergone Landau damping, acquiring a long tail with several small Stoner features. Despite the damping, the magnon peak just barely remains coherent in the sense that $\mathrm{Re}\:\xi^{++}_0(\omega)$ crosses unity at the peak position, in contrast to e.g. the $n=2$ majority mode at the K-point. 
For the minority magnon mode at $\mathbf{q}_\Gamma$, the picture is, once again, very similar. A single well-defined peak in the Stoner spectrum %(due in part to the almost flat minority $d$-band below the Fermi level) 
creates a pole-like feature in $\mathrm{Re}\:\xi^{--}_0(\omega)$ (via the Kramers-Kronig relation) resulting in the collective enhancement of a magnon peak at the lower end of the continuum. Seemingly, the minority Stoner spectrum lacks sufficient amplitude to generate a coherent minority resonance, but as we will show below, the resonance can actually be recategorized as coherent once corrected for the artifical broadening applied in the ALDA calculations. Crucially, one may also notice that $\mathrm{Re}\:\xi^{--}_0(\omega)$ does not proceed to vanish below the Stoner continuum, but increases instead thanks to the overlap with the $n=2$ majority mode. This implies that the majority Stoner continuum contributes constructively to the minority magnon enhancement (and vice-versa), which is essential in order to explain the existence of a coherent minority magnon despite the limited density of unoccupied majority $d$-states in Fe$_3$GeTe$_2$, see Fig. \ref{fig:Fe3GeTe2_crystalstructure_bandstructure_and_pdos}. 
%In fact, had the low energy minority mode overlapped with the acoustic $n=0$ majority mode, for which 
%$\mathrm{Re}\:\xi^{++}_0(\mathbf{q}_\Gamma,\omega=0)=1$,
%the minority Stoner continuum depicted in Fig. \ref{fig:Fe3GeTe2_enhancement_analysis} would indeed have been intense enough to produce a coherent minority resonance. %As a result, we find it extremely likely that coherent minority magnons will be observed in the future for ferromagnets comparable to Fe$_3$GeTe$_2$ in many qualitative aspects. 
For the optical majority magnon modes, we identify in Fig. \ref{fig:Fe3GeTe2_enhancement_analysis} two main reasons for the mode branching (multitude of peaks) observed in the magnon lineshapes. The first is due to Stoner satellites (in Ni referred to as Stoner stripes \cite{Friedrich2020}), here understood as weak but well-defined peaks in the Stoner spectrum below the main continuum. For the $n=1$ mode at $\mathbf{q}_\mathrm{K}$, such a Stoner satellite is placed right around the main magnon resonance, generating an extra wiggle in $\mathrm{Re}\:\xi^{++}_1(\omega)$ and a corresponding extra peak in the magnon lineshape. Extra branches in the magnon dispersion of this sort, we refer to as satellite magnons. The second is due to cases where the main Stoner continuum has multiple peaks, such as for the optical $n=2$ majority mode at $\mathbf{q}_\Gamma$. In this scenario, a magnon branch can reside in the valley in between the two Stoner peaks, forming a second coherent magnon resonance. Such magnon branches we refer to as valley magnons, and due to Landau damping we often find that the lineshape is peaked further towards the bottom of the Stoner valley than what the real part of the self-enhancement function would suggest.

\subsection{Minority magnon band structure}

\begin{figure}[t]
    \centering
    \includegraphics[scale=1.0]{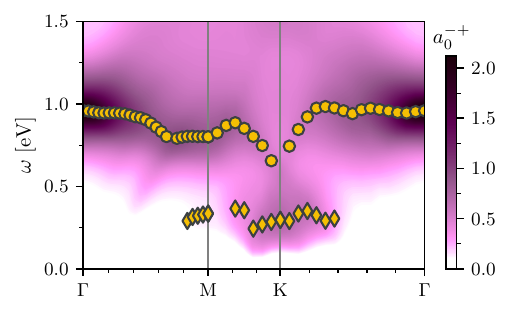}
    \caption{Magnon band structure for the low energy minority magnon mode. The magnon mode lineshape $a_0^{-+}(\mathbf{q},\omega)$ is shown as a colored contour on top of which the peak positions of the identified magnon branches are shown. The main minority branch is marked with disks and the spin-inversion magnon branch with rhombuses.}
    \label{fig:Fe3GeTe2_minority_magnon_bandstructure}
\end{figure}
%
%Following the procedure outlined above, 
Following this analysis, we have computed the full scattering function as a function of wave vector $\mathbf{q}$, extracted the majority and minority magnon lineshapes and analyzed the self-enhancement function in order to categorize each of the magnon peaks. 
In Fig. \ref{fig:Fe3GeTe2_minority_magnon_bandstructure}, we present the %calculated ALDA magnon branches of the $n=0$ minority mode of monolayer Fe$_3$GeTe$_2$. 
resulting magnon band structure of the $n=0$ minority magnon mode.
%For the acoustic Goldstone mode, we show the raw ALDA dispersion with an observed gap error of 61 meV. Normally, one would correct this error by applying a rigid shift to the frequency axis \cite{Skovhus2021,Singh2018} or by using one of a series of more comprehensive correction schemes \cite{Buczek2011b,Lounis2011,Rousseau2012}. Regardless, we have not been able to converge the acoustic magnon dispersion within current computational constraints, why we choose to focus on the converged minority and optical majority magnon modes instead. It should, however, be mentioned in passing, that with a plane-wave cutoff of 500 eV as shown in Fig. \ref{fig:Fe3GeTe2_magnon_bandstructure},  the ALDA magnon dispersion suggests the ferromagnetic state to be dynamically unstable, since the minimum of the dispersion lies at the K-point rather than at the $\Gamma$-point. 
We see that the main branch attains its largest scattering amplitude close to the $\Gamma$-point, %where the resonance is near-coherent, 
while it gets progressively damped as the corresponding Stoner peak is broadened for finite $\mathbf{q}$. Near the BZ boundary, where wave vectors $\mathbf{q}$ connect the majority and minority Fermi surfaces, the branch is further damped from below due to spin-inverted majority Stoner pairs \cite{SkovhusPhD}. These are the pairs that give rise to the conventional minority scattering observed in the homogeneous electron gas \cite{Friedrich2020}, and at the K-point the corresponding spin-inversion magnon completely dominates the low energy spectrum to which the main branch becomes a shoulder in the lineshape, see Fig. \ref{fig:Fe3GeTe2_enhancement_analysis}. In Appendix \ref{sec:majority band structures}, we provide analogous band structure figures for the majority magnon modes and compare our results in this regard to a previous RPA study \cite{Costa2020a} of the majority modes, as well as the spin-wave dispersion within the Heisenberg model calculated using magnetic force theorem exchange parameters \cite{Liechtenstein1987,Bruno2003,Durhuus2023}.

\subsection{Minority magnon enhancement---the full picture}

\begin{figure*}[tb]
    \centering
    \includegraphics[scale=1.0]{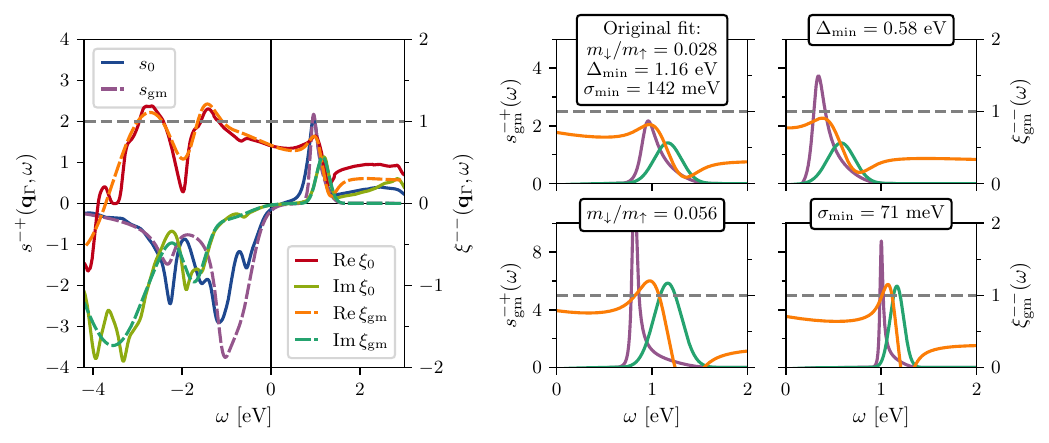}
    \caption{Minority magnon enhancement in the long wavelength limit. The minority self-enhancement function $\xi^{--}$ of the low energy minority mode (right axes) gives rise to a minority magnon peak in the corresponding scattering function $s^{-+}$ at positive frequencies (left axes) as $\xi^{--}(\omega)\rightarrow 1$. If $\mathrm{Re}\:\xi^{--}$ crosses unity, the resulting minority magnon is coherent. To the left, the ALDA self-enhancement and scattering functions $\xi^{--}_0$ and $s^{-+}_0(\omega) = a^{-+}_0(\omega) - a^{+-}_2(-\omega)$ are compared to a scalar gaussian model fitted to the Stoner peaks in $\mathrm{Im}\:\xi^{--}_0(\omega)$, with a model scattering function calculated based on Eq. \eqref{eq:gaussian model}. To the right, the minority side of the gaussian fit is shown in more detail, illustrating the increased minority magnon enhancement that happens if the relative minority weight $m_\downarrow/m_\uparrow$ is doubled or either the minority Stoner peak frequency $\Delta_\mathrm{min}$ or width $\sigma_\mathrm{min}$ are halved.}
    \label{fig:Fe3GeTe2_minority_magnon_modelling}
\end{figure*}
For the remaining main body of text, we focus on the minority magnon enhancement in order to fully address the coherency question concerning the minority magnons in Fe$_3$GeTe$_2$. To do so, we show in Fig. \ref{fig:Fe3GeTe2_minority_magnon_modelling} the full self-enhancement function of the $n=0$ minority magnon mode along with the corresponding scattering function, which at negative frequencies is given by the $n=2$ majority magnon mode. Focusing on $\mathrm{Re}\:\xi^{--}_0(\omega)$, we see how both the minority and majority Stoner pairs add positive weight to $\mathrm{Re}\:\xi$ for frequencies in between the respective Stoner continua (represented by $\mathrm{Im}\:\xi^{--}_0(\omega)$) and thus contribute constructively to the magnon enhancement in this region. Had the majority Stoner pairs not contributed constructively to the magnon enhancement at the minority side of the frequency axis, the pole-like feature in $\mathrm{Re}\:\xi_0^{--}(\omega)$ around $\omega=1$ eV would have been centered around $\mathrm{Re}\:\xi^{--}\sim 0$, not $\sim 0.5$, and the minority magnon would be far from coherency. Based on the imaginary part of the self-enhancement function, we introduce in Fig. \ref{fig:Fe3GeTe2_minority_magnon_modelling} also a simplified model for the magnon enhancement in this mode. Namely, we fit four gaussians (corresponding to one peak at the minority side and three on the majority side) to $\mathrm{Im}\:\xi^{--}_0(\omega)$ and use the Kramers-Kronig relation to obtain the real part of our gaussian model $\xi^{--}_\mathrm{gm}(\omega)$. In order to make our model $\mathrm{Re}\:\xi^{--}_\mathrm{gm}(\omega)$ reproduce $\mathrm{Re}\:\xi^{--}_0(\omega)$ as closely as possible in between the main majority and minority Stoner peaks, we have also added an additional majority Stoner peak outside the calculated frequency range at $\omega=-6.4\:\mathrm{eV}$. We can then calculate a model scattering function by inverting the scalar Dyson equation
\begin{equation}
    \chi^{-+}_\mathrm{gm}(\omega) = \frac{\chi^{-+}_\mathrm{KS,gm}(\omega)}{1 - \xi^{--}_\mathrm{gm}(\omega)},
    \label{eq:gaussian model}
\end{equation}
where $\xi^{--}_\mathrm{gm}(\omega) = \chi^{-+}_\mathrm{KS,gm}(\omega) f_\mathrm{xc}^{+-}$. For modelling purposes, we treat $f_\mathrm{xc}^{+-}$ as a free parameter, fixing it so that the peak intensity of the minority magnon is reproduced, a choice which also results in a qualitatively reasonable description of the majority magnons, see Fig. \ref{fig:Fe3GeTe2_minority_magnon_modelling}. Central to the model, we highlight here three key parameters for the minority magnon enhancement: the weight of the minority Stoner peak relative to the sum of majority weights $m_\downarrow/m_\uparrow$, the central frequency of the minority Stoner peak $\Delta_\mathrm{min}$ (can be thought of as a minority spin splitting) and the width of the minority Stoner peak (here given in terms of the standard deviation of the gaussian fit $\sigma_\mathrm{min}$). Whereas $\mathrm{Re}\:\xi^{--}_0(\omega)$ and $\mathrm{Re}\:\xi^{--}_\mathrm{gm}(\omega)$ only \textit{approach} unity at positive frequencies, one can then ask: what happens, if one or more of the key parameters change? For example, if we add an equal amount of minority and majority Stoner weight (in the latter case distributed among the gaussian fits relative to their fitted weights), we change the $m_\downarrow/m_\uparrow$ fraction while leaving the total weight $-m=m_\downarrow-m_\uparrow$ unaltered. Since the total weight $m$ is proportional to the spin-polarization of the ground state thanks to a sum rule \cite{Kubo1966,Jensen1991,Rousseau2012,Skovhus2021}, such a change should not affect $f_\mathrm{xc}^{+-}$. In Fig. \ref{fig:Fe3GeTe2_minority_magnon_modelling} we show that doubling $m_\downarrow/m_\uparrow$ results in a clear coherent magnon resonance. Similarly, we show that the minority magnon gets closer to coherency if $\Delta_\mathrm{min}$ is decreased (increasing the enhancement gained from the majority Stoner pairs) and that minority magnon coherency is obtained also if the width of the minority Stoner peak is halved (decreasing the damping of the pole-like feature in $\mathrm{Re}\:\xi^{--}_\mathrm{gm}(\omega)$). 
\begin{figure}[tb]
    \centering
    \includegraphics[scale=1.0]{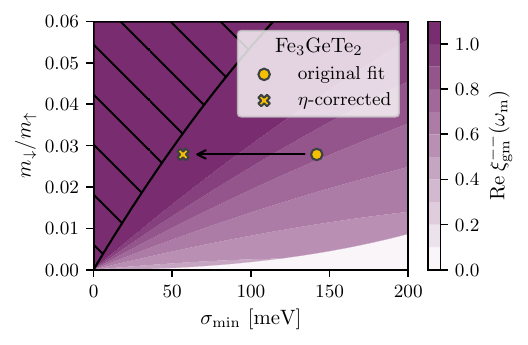}
    \caption{Contour map of the modelled minority magnon coherency, quantified in terms of $\mathrm{Re}\:\xi_\mathrm{gm}^{--}(\omega_\mathrm{m})$ where $\omega_\mathrm{m}=\mathrm{argmin}_{0<\omega<\Delta_\mathrm{min}}{|1 - \mathrm{Re}\:\xi_\mathrm{gm}^{--}(\omega)|}$, as a function of the relative weight of the minority Stoner peak $m_\downarrow/m_\uparrow$ and its width $\sigma_\mathrm{min}$. The hatched area indicates parameters for which a coherent minority magnon resonance lies more than $3\sigma_\mathrm{min}$ below $\Delta_\mathrm{min}$.}
    \label{fig:Fe3GeTe2_minority_magnon_diagram}
\end{figure}
The latter of these scenarios is particularly relevant in order to understand the full implications of the presently presented ALDA results for monolayer Fe$_3$GeTe$_2$. When calculating $\chi_\mathrm{KS}^{+-}$, an artificial spectral broadening of $\eta=100\:\mathrm{meV}$ was applied to achieve $k$-point convergence of the magnon dispersion. However, this also broadens the resulting spectra, adding $2\eta$ of extra width to all spectral features. In Fig. \ref{fig:Fe3GeTe2_minority_magnon_diagram}, we map out the minority magnon coherency in the gaussian model as a function of $m_\downarrow/m_\uparrow$ and $\sigma_\mathrm{min}$. If one subtracts the artificial broadening $\eta$ from the half width at half maximum of the Stoner peak, $\mathrm{HWHM}=\sqrt{2\ln{2}}\sigma_\mathrm{min}$, the low energy minority magnon mode of monolayer Fe$_3$GeTe$_2$ is categorized as coherent, despite the fact that the minority Stoner weight only amounts to $2.8\%$ of the majority weight in the mode. Furthermore, Landau damping is expected to be weak, and the minority magnon peak should in reality look more like the sharp spectral feature to the bottom right of Fig. \ref{fig:Fe3GeTe2_minority_magnon_modelling} than the bare output of our ALDA calculations.

%%%%% Conclusion and outlook %%%%%
% Here we make our final remarks

\section{Conclusion and outlook}

To summarize, we have showcased how minority magnon excitations enter the first principles theory of linear response and shown that monolayer Fe$_3$GeTe$_2$ hosts three such collective modes in the ALDA. Along with the pronounced mode branching of the optical majority modes, we have studied the lowest energy minority magnon mode in detail. Despite the low density of unoccupied majority Fe $d$-states, we do expect the mode to constitute a coherent minority magnon resonance close to the $\Gamma$-point, in particular thanks to the flat occupied minority $d$-band in the vicinity of the Fermi level. %. Searching in the future for ferromagnets with at least one fully occupied minority $d$-band and a more extensive unoccupied density of majority $d$-states than in the present case, we believe that the discovery of ferromagnets with coherent minority magnon modes is eminent.
%Based on these findings, we believe the experimental discovery of minority magnons in Fe$_3$GeTe$_2$ or similar nontrivial ferromagnets to be eminent.
In LDA, this band lies 0.71--0.83 eV below the Fermi level, but with an improved description of local correlations the band seems to move even closer, lying just 0.2--0.3 eV below it at the DMFT level \cite{Ghosh2023}. If this is indeed the case, the minority Stoner gap should in reality be smaller than what is found here using ALDA, which will increase the minority magnon enhancement thanks to the increased spectral proximity of majority magnons, and significantly redshift the minority magnon frequency.
For this reason, we find it more likely to actually observe minority magnons in monolayer Fe$_3$GeTe$_2$ within the 0.25--0.5 eV range, instead of the 0.8--1.0 eV range that the bare ALDA results suggest. 

Based on the work presented above, and the generality of the underlying physics, we believe that the experimental discovery of minority magnons in a ferromagnet is impending, for example with techniques such as SPEELS \cite{Zakeri2014,Qin2015}. The key ingredient is an occupied minority $d$-band in combination with vacant majority $d$-states above the Fermi level (both preferably with a narrow density of states), of which monolayer Fe$_3$GeTe$_2$ is a promising candidate, as well as the wider Fe$_x$GeTe$_2$ family \cite{Wang2023a}.

\appendix

\section{Computational details}\label{sec:computational details}

\begin{figure}[b]
    \centering
    \includegraphics{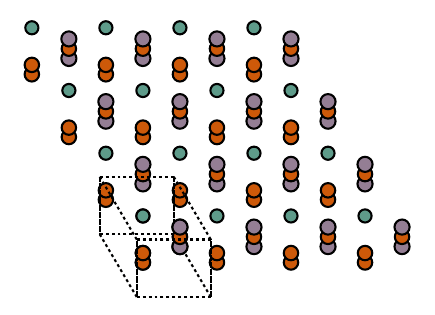}
    \caption{Top view of the Fe$_3$GeTe$_2$ monolayer crystal. For a side view, see Fig. \ref{fig:Fe3GeTe2_crystalstructure_bandstructure_and_pdos}.}
    \label{fig:Fe3GeTe2_crystal_structure}
\end{figure}
Studying the transverse magnetic excitations of monolayer Fe$_3$GeTe$_2$, we have used the relaxed crystal structure from the Computational 2D Materials Database C2DB \cite{Haastrup2018,Gjerding2021}, adjusting the vacuum to 10 $\mathrm{\AA}$ (see also Fig. \ref{fig:Fe3GeTe2_crystal_structure}). We calculate the LDA ground state on a $48 \times 48 \times 1$ $\Gamma$-centered $k$-point grid, based on which $\chi^{+-}_\mathrm{KS}$ and $\Xi_\mathrm{ALDA}^{++}$ are calculated using 12 empty-shell bands per atom, a plane-wave cutoff of 500 eV and a spectral broadening of 100 meV. Inverting the Dyson equation, we obtain $\chi^{+-}$, from which the scattering function $S^{+-}$ is extracted via its anti-Hermitian part in the plane-wave components.

Whereas Fig. \ref{fig:Fe3GeTe2_total_scattering} shows the trace of $S^{+-}$ and $S_\mathrm{KS}^{+-}$ in the full 500 eV plane-wave basis, we subsequently reduce $S^{+-}$ to a plane-wave cutoff of 200 eV before diagonalizing it to extract $A^{+-}$ and $A^{-+}$. We save the full 200 eV trace of $A^{+-}$ and $A^{-+}$ (to later construct the corresponding many-body Stoner spectra), but only the 27 and 15 largest actual eigenvalues and corresponding eigenvectors respectively.
%In order to analyze the magnon modes independently, see e.g. Fig. 2 of the main text, we reduce $S^{+-}$ to a plane-wave cutoff of 200 eV, diagonalize it and save the 27 and 15 largest eigenvalues of $A^{+-}$ and $A^{-+}$ respectively, along with the corresponding eigenvectors and full eigenvalue trace. 
This basis reduction is basically a disk space saving exercise due to the need of running calculations across many different jobs. It should be noted that we have diagonalized $S_\mathrm{KS}^{+-}$ in the 500 eV plane-wave basis and show in Fig. \ref{fig:Fe3GeTe2_mode_extraction} the 500 eV trace of $A_\mathrm{KS}^{+-}$ and $A_\mathrm{KS}^{-+}$, but rescaled to the number of basis functions in the 200 eV basis in order to provide a fair comparison to the many-body Stoner spectra. We also save an eigendecomposition of the self-enhancement function consisting of its 42 largest eigenvalues, such that we can reconstruct both $A^{+-}(\mathbf{q},\omega)$, $A^{-+}(\mathbf{q},\omega)$ and $\Xi^{++}(\mathbf{q},\omega)$ after all calculations have finished. %Apart from $S^{+-}$, $S_\mathrm{KS}^{+-}$, $A_\mathrm{KS}^{+-}$ and $A_\mathrm{KS}^{-+}$ in Figs. 1 and 2 of main text (which are shown in the full 500 eV cutoff basis) all results of the letter are based on the reduced eigendecompositions in the 200 eV basis.
From the spectral eigendecompositions, we extract the magnon lineshapes as the inner product $a_n(\mathbf{q},\omega) = \langle v_n| A(\mathbf{q},\omega) |v_n\rangle$, using magnon eigenvectors $|v_n\rangle$ taken 
at a single frequency $\omega$. For the majority magnon modes, we choose $\omega$ to maximize
the minimum eigenvalue difference between the three modes: $\max_\omega \left( \min_{n>0} \left[\lambda_{n-1}(\omega) - \lambda_n(\omega)\right]\right)$. 
This ensures the extraction of well-defined lineshapes, since the eigenvectors are only sensitive to the frequency in the vicinity of crossing eigenvalues.
For the minority magnons, we focus on the lowest energy mode, extracting its mode vector at the eigenvalue maximum. The only exception is Fig. \ref{fig:Fe3GeTe2_mode_extraction}, where we showcase that also the three minority magnon mode lineshapes can be extracted in a similar fashion to the majority modes, albeit a little less convincingly since the $n=2$ mode does not extend to lower frequencies than 2 eV. 

\section{Majority magnon band structures}\label{sec:majority band structures}

\subsection{Majority magnons in ALDA}
\begin{figure}[b]
    \centering
    \includegraphics{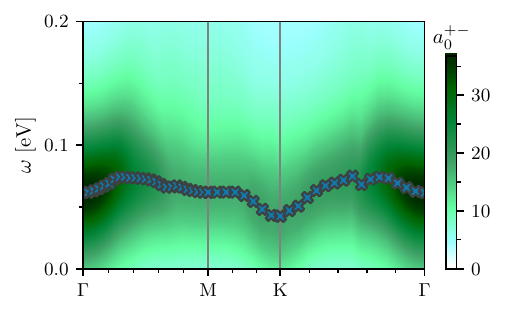}
    \caption{Magnon band structure for the acoustic majority mode. The lineshape $a_0^{+-}(\mathbf{q},\omega)$ (color contour) was calculated with a reduced broadening of 50 meV, and the peak positions of the coherent Goldstone branch are shown with crosses.}
    \label{fig:Fe3GeTe2_acoustic_magnon_bandstructure}
\end{figure}
\begin{figure*}[tb]
    \centering
    \includegraphics[scale=1.0]{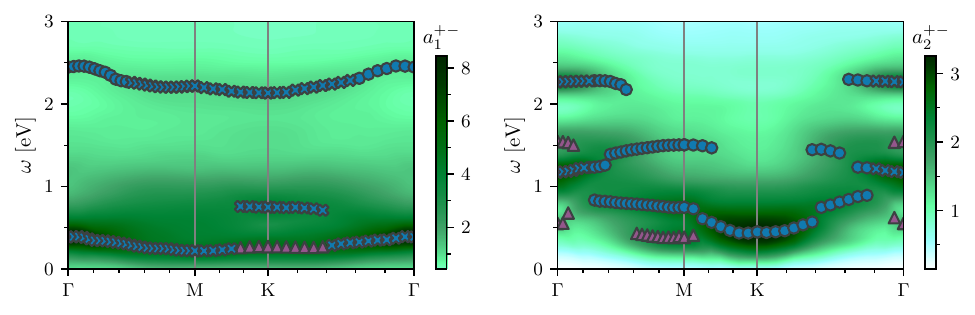}
    \caption{Optical majority magnon band structure. The magnon mode lineshapes $a_n^{+-}(\mathbf{q},\omega)$ are shown as colored contour maps on top of which the peak positions of the identified magnon branches are shown. Peaks corresponding to coherent magnon resonances (as extracted from the bare ALDA calculations) are marked with crosses, incoherent magnon resonances with disks and satellite magnons (which are all incoherent) with triangles.}
    \label{fig:Fe3GeTe2_optical_magnon_bandstructure}
\end{figure*}
In Figs. \ref{fig:Fe3GeTe2_acoustic_magnon_bandstructure} and \ref{fig:Fe3GeTe2_optical_magnon_bandstructure}, we present the ALDA magnon band structures of the majority modes in monolayer Fe$_3$GeTe$_2$.
For the acoustic Goldstone mode, we show the raw ALDA dispersion with an observed gap error of 61 meV (due to the imperfect numerical convergence of this mode). Normally, one would correct this error by applying a rigid shift to the frequency axis \cite{Skovhus2021,Singh2018} or by using one of a series of more comprehensive correction schemes \cite{Buczek2011b,Lounis2011,Rousseau2012}. Regardless, we have not been able to converge the acoustic magnon dispersion within current computational constraints, why we have devoted our focus to the converged minority and optical majority magnon modes instead. It should, however, be mentioned in passing, that with a plane-wave cutoff of 500 eV as shown in Fig. \ref{fig:Fe3GeTe2_acoustic_magnon_bandstructure}, the ALDA magnon dispersion suggests the ferromagnetic state to be dynamically unstable, since the minimum of the dispersion lies at the K-point rather than at the $\Gamma$-point.
Continuing to the $n=1$ optical mode, see Fig. \ref{fig:Fe3GeTe2_optical_magnon_bandstructure}, there are two main branches at the $\Gamma$-point: a conventional coherent magnon branch around 390 meV and an incoherent valley magnon around 2.45 eV. Approaching the BZ edge, the valley magnon becomes coherent, whereas the conventional optical magnon branch splits in two, of which the lower one is categorized as a satellite magnon. In a previous RPA study of the magnon spectrum in monolayer Fe$_3$GeTe$_2$ \cite{Costa2020a}, the $n=1$ optical mode has been characterized as "nonbonding" (in a sublattice sense) since it only carries weight on the two Fe(I) atoms, see Fig. \ref{fig:Fe3GeTe2_mode_extraction}. In that study, only the lower magnon branch was extracted (following what we refer to as the satellite magnon), and the existence of a coherent high-frequency magnon branch was overlooked. Where they find a relatively flat magnon dispersion with a bandwidth of $\sim50$ meV, we find a bandwidth (following the satellite magnon) of 175 meV, in both cases dispersing to lower frequencies away from the $\Gamma$-point and centered around 300 meV. Lastly, for the $n=2$ optical magnon mode, we find two coherent branches at the $\Gamma$-point; a conventional magnon branch and a valley magnon, with satellite magnons below and in between, see also Fig. \ref{fig:Fe3GeTe2_enhancement_analysis}. In the preceding RPA study \cite{Costa2020a}, which is limited to frequencies up to 1 eV, no coherent branches corresponding to the $n=2$ mode are reported on. As the coherent branches disperse to higher frequencies and decohere away from the $\Gamma$-point, a new incoherent main branch appears at lower frequencies, with a satellite magnon branch at its foot around the M-point. Although it is not the focus of the RPA study \cite{Costa2020a}, this lower main branch is clearly present in their data, but seemingly with a slightly smaller bandwidth than the 460 meV we find here.

\subsection{Magnons in the Heisenberg model}

\begin{figure}[tb]
    \centering
    \includegraphics[scale=1.0]{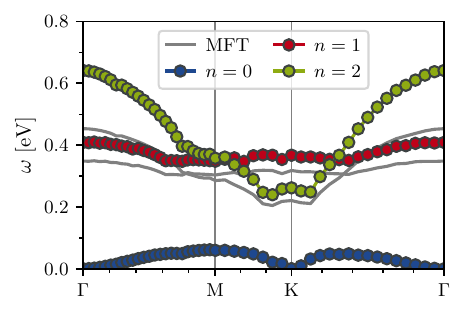}
    \caption{Spin-wave dispersion obtained from linear response MFT. Numbered and in colors, we show the results obtained with the renormalization scheme of Bruno \cite{Bruno2003}, in grey the bare MFT results obtained from the original approach \cite{Liechtenstein1987,Durhuus2023}.}
    \label{fig:Fe3GeTe2_MFT}
\end{figure}
In Fig. \ref{fig:Fe3GeTe2_MFT} we show the spin-wave band structure obtained from linear-response magnetic force theorem (MFT) calculations \cite{Liechtenstein1987,Bruno2003,Durhuus2023} using the same computational parameters as in the ALDA calculations. This approach enforces a Heisenberg description upon the magnetic structure and yields three distinct solutions to the spin-wave equation of motion, one for each Fe atom in the unit cell. We find that the three spin-wave modes (as obtained from diagonalizing the three-site dynamical matrix) match well the ALDA eigenmodes shown in Fig. \ref{fig:Fe3GeTe2_mode_extraction}, yielding a dispersion loosely reminiscent of the lowest ALDA magnon branch for each of the three modes. Compared to the rich mode branching observed in the ALDA, it is clear that the Heisenberg modelling approach fails rather dramatically in providing a comprehensive description of the optical magnetic excitations. This is expected, since the rather complicated ALDA lineshapes emerge due to the coupling to the Stoner continuum, which is not explicitly accounted for in the localized Heisenberg spin model. For example, one may notice that the $n=1$ optical mode does not indicate any form of branching in the vicinity of the $\mathrm{K}$-point and significantly underestimates the bandwidth (65 meV) in comparison to ALDA. For the $n=2$ mode, the MFT dispersion seems to mimic the incoherent low energy magnon branch throughout most of the Brillouin zone, but again the magnon energy as well as the bandwidth (400 meV) are underestimated in comparison to ALDA. %We do note, however, that the three spin-wave modes (as obtained from diagonalizing the three-site dynamical matrix) match the eigenmodes shown in Fig. \ref{fig:Fe3GeTe2_mode_extraction}. %This is probably due to the strong Landau damping, which is not accounted for in the Heisenberg description.

\bibliography{bibliography}

\begin{thebibliography}{39}%
\makeatletter
\providecommand \@ifxundefined [1]{%
 \@ifx{#1\undefined}
}%
\providecommand \@ifnum [1]{%
 \ifnum #1\expandafter \@firstoftwo
 \else \expandafter \@secondoftwo
 \fi
}%
\providecommand \@ifx [1]{%
 \ifx #1\expandafter \@firstoftwo
 \else \expandafter \@secondoftwo
 \fi
}%
\providecommand \natexlab [1]{#1}%
\providecommand \enquote  [1]{``#1''}%
\providecommand \bibnamefont  [1]{#1}%
\providecommand \bibfnamefont [1]{#1}%
\providecommand \citenamefont [1]{#1}%
\providecommand \href@noop [0]{\@secondoftwo}%
\providecommand \href [0]{\begingroup \@sanitize@url \@href}%
\providecommand \@href[1]{\@@startlink{#1}\@@href}%
\providecommand \@@href[1]{\endgroup#1\@@endlink}%
\providecommand \@sanitize@url [0]{\catcode `\\12\catcode `\$12\catcode `\&12\catcode `\#12\catcode `\^12\catcode `\_12\catcode `\%12\relax}%
\providecommand \@@startlink[1]{}%
\providecommand \@@endlink[0]{}%
\providecommand \url  [0]{\begingroup\@sanitize@url \@url }%
\providecommand \@url [1]{\endgroup\@href {#1}{\urlprefix }}%
\providecommand \urlprefix  [0]{URL }%
\providecommand \Eprint [0]{\href }%
\providecommand \doibase [0]{https://doi.org/}%
\providecommand \selectlanguage [0]{\@gobble}%
\providecommand \bibinfo  [0]{\@secondoftwo}%
\providecommand \bibfield  [0]{\@secondoftwo}%
\providecommand \translation [1]{[#1]}%
\providecommand \BibitemOpen [0]{}%
\providecommand \bibitemStop [0]{}%
\providecommand \bibitemNoStop [0]{.\EOS\space}%
\providecommand \EOS [0]{\spacefactor3000\relax}%
\providecommand \BibitemShut  [1]{\csname bibitem#1\endcsname}%
\let\auto@bib@innerbib\@empty
%</preamble>
\bibitem [{\citenamefont {Yosida}(1996)}]{Yosida1996}%
  \BibitemOpen
  \bibfield  {author} {\bibinfo {author} {\bibfnamefont {K.}~\bibnamefont {Yosida}},\ }\href@noop {} {\emph {\bibinfo {title} {Theory of Magnetism}}},\ \bibinfo {series} {Springer Series in Solid-State Sciences}, Vol.\ \bibinfo {volume} {122}\ (\bibinfo  {publisher} {Springer-Verlag Berlin Heidelberg},\ \bibinfo {year} {1996})\BibitemShut {NoStop}%
\bibitem [{\citenamefont {Szilva}\ \emph {et~al.}(2023)\citenamefont {Szilva}, \citenamefont {Kvashnin}, \citenamefont {Stepanov}, \citenamefont {Nordstr\"om}, \citenamefont {Eriksson}, \citenamefont {Lichtenstein},\ and\ \citenamefont {Katsnelson}}]{Szilva2023}%
  \BibitemOpen
  \bibfield  {author} {\bibinfo {author} {\bibfnamefont {A.}~\bibnamefont {Szilva}}, \bibinfo {author} {\bibfnamefont {Y.}~\bibnamefont {Kvashnin}}, \bibinfo {author} {\bibfnamefont {E.~A.}\ \bibnamefont {Stepanov}}, \bibinfo {author} {\bibfnamefont {L.}~\bibnamefont {Nordstr\"om}}, \bibinfo {author} {\bibfnamefont {O.}~\bibnamefont {Eriksson}}, \bibinfo {author} {\bibfnamefont {A.~I.}\ \bibnamefont {Lichtenstein}},\ and\ \bibinfo {author} {\bibfnamefont {M.~I.}\ \bibnamefont {Katsnelson}},\ }\href {https://doi.org/10.1103/RevModPhys.95.035004} {\bibfield  {journal} {\bibinfo  {journal} {Rev. Mod. Phys.}\ }\textbf {\bibinfo {volume} {95}},\ \bibinfo {pages} {035004} (\bibinfo {year} {2023})}\BibitemShut {NoStop}%
\bibitem [{\citenamefont {Squires}(1978)}]{Squires1978}%
  \BibitemOpen
  \bibfield  {author} {\bibinfo {author} {\bibfnamefont {G.~L.}\ \bibnamefont {Squires}},\ }\href@noop {} {\emph {\bibinfo {title} {Thermal Neutron Scattering}}}\ (\bibinfo  {publisher} {Cambridge University Press},\ \bibinfo {year} {1978})\BibitemShut {NoStop}%
\bibitem [{\citenamefont {Jensen}\ and\ \citenamefont {Mackintosh}(1991)}]{Jensen1991}%
  \BibitemOpen
  \bibfield  {author} {\bibinfo {author} {\bibfnamefont {J.}~\bibnamefont {Jensen}}\ and\ \bibinfo {author} {\bibfnamefont {A.~R.}\ \bibnamefont {Mackintosh}},\ }\href@noop {} {\emph {\bibinfo {title} {Rare Earth Magnetism: Structures and excitations}}},\ The International Series of Monographs on Physics\ (\bibinfo  {publisher} {Clarendon Press, Oxford},\ \bibinfo {year} {1991})\BibitemShut {NoStop}%
\bibitem [{\citenamefont {Boothroyd}(2020)}]{Boothroyd2020}%
  \BibitemOpen
  \bibfield  {author} {\bibinfo {author} {\bibfnamefont {A.~T.}\ \bibnamefont {Boothroyd}},\ }\href@noop {} {\emph {\bibinfo {title} {Principles of Neutron Scattering from Condensed Matter}}}\ (\bibinfo  {publisher} {Oxford University Press},\ \bibinfo {year} {2020})\BibitemShut {NoStop}%
\bibitem [{\citenamefont {Savrasov}(1998)}]{Savrasov1998}%
  \BibitemOpen
  \bibfield  {author} {\bibinfo {author} {\bibfnamefont {S.~Y.}\ \bibnamefont {Savrasov}},\ }\href {https://doi.org/10.1103/PhysRevLett.81.2570} {\bibfield  {journal} {\bibinfo  {journal} {Physical Review Letters}\ }\textbf {\bibinfo {volume} {81}},\ \bibinfo {pages} {2570} (\bibinfo {year} {1998})}\BibitemShut {NoStop}%
\bibitem [{\citenamefont {Karlsson}\ and\ \citenamefont {Aryasetiawan}(2000)}]{Karlsson2000}%
  \BibitemOpen
  \bibfield  {author} {\bibinfo {author} {\bibfnamefont {K.}~\bibnamefont {Karlsson}}\ and\ \bibinfo {author} {\bibfnamefont {F.}~\bibnamefont {Aryasetiawan}},\ }\href {https://doi.org/10.1103/PhysRevB.62.3006} {\bibfield  {journal} {\bibinfo  {journal} {Physical Review B}\ }\textbf {\bibinfo {volume} {62}},\ \bibinfo {pages} {3006} (\bibinfo {year} {2000})}\BibitemShut {NoStop}%
\bibitem [{\citenamefont {Buczek}\ \emph {et~al.}(2011)\citenamefont {Buczek}, \citenamefont {Ernst},\ and\ \citenamefont {Sandratskii}}]{Buczek2011b}%
  \BibitemOpen
  \bibfield  {author} {\bibinfo {author} {\bibfnamefont {P.}~\bibnamefont {Buczek}}, \bibinfo {author} {\bibfnamefont {A.}~\bibnamefont {Ernst}},\ and\ \bibinfo {author} {\bibfnamefont {L.~M.}\ \bibnamefont {Sandratskii}},\ }\href {https://doi.org/10.1103/PhysRevB.84.174418} {\bibfield  {journal} {\bibinfo  {journal} {Physical Review B}\ }\textbf {\bibinfo {volume} {84}},\ \bibinfo {pages} {174418} (\bibinfo {year} {2011})}\BibitemShut {NoStop}%
\bibitem [{\citenamefont {Rousseau}\ \emph {et~al.}(2012)\citenamefont {Rousseau}, \citenamefont {Eiguren},\ and\ \citenamefont {Bergara}}]{Rousseau2012}%
  \BibitemOpen
  \bibfield  {author} {\bibinfo {author} {\bibfnamefont {B.}~\bibnamefont {Rousseau}}, \bibinfo {author} {\bibfnamefont {A.}~\bibnamefont {Eiguren}},\ and\ \bibinfo {author} {\bibfnamefont {A.}~\bibnamefont {Bergara}},\ }\href {https://doi.org/10.1103/PhysRevB.85.054305} {\bibfield  {journal} {\bibinfo  {journal} {Physical Review B}\ }\textbf {\bibinfo {volume} {85}},\ \bibinfo {pages} {054305} (\bibinfo {year} {2012})}\BibitemShut {NoStop}%
\bibitem [{\citenamefont {Cao}\ \emph {et~al.}(2018)\citenamefont {Cao}, \citenamefont {Lambert}, \citenamefont {Radaelli},\ and\ \citenamefont {Giustino}}]{Cao2017}%
  \BibitemOpen
  \bibfield  {author} {\bibinfo {author} {\bibfnamefont {K.}~\bibnamefont {Cao}}, \bibinfo {author} {\bibfnamefont {H.}~\bibnamefont {Lambert}}, \bibinfo {author} {\bibfnamefont {P.~G.}\ \bibnamefont {Radaelli}},\ and\ \bibinfo {author} {\bibfnamefont {F.}~\bibnamefont {Giustino}},\ }\href {https://doi.org/10.1103/PhysRevB.97.024420} {\bibfield  {journal} {\bibinfo  {journal} {Physical Review B}\ }\textbf {\bibinfo {volume} {97}},\ \bibinfo {pages} {024420} (\bibinfo {year} {2018})}\BibitemShut {NoStop}%
\bibitem [{\citenamefont {Singh}\ \emph {et~al.}(2019)\citenamefont {Singh}, \citenamefont {Elliott}, \citenamefont {Nautiyal}, \citenamefont {Dewhurst},\ and\ \citenamefont {Sharma}}]{Singh2018}%
  \BibitemOpen
  \bibfield  {author} {\bibinfo {author} {\bibfnamefont {N.}~\bibnamefont {Singh}}, \bibinfo {author} {\bibfnamefont {P.}~\bibnamefont {Elliott}}, \bibinfo {author} {\bibfnamefont {T.}~\bibnamefont {Nautiyal}}, \bibinfo {author} {\bibfnamefont {J.~K.}\ \bibnamefont {Dewhurst}},\ and\ \bibinfo {author} {\bibfnamefont {S.}~\bibnamefont {Sharma}},\ }\href {https://doi.org/10.1103/PhysRevB.99.035151} {\bibfield  {journal} {\bibinfo  {journal} {Physical Review B}\ }\textbf {\bibinfo {volume} {99}},\ \bibinfo {pages} {035151} (\bibinfo {year} {2019})}\BibitemShut {NoStop}%
\bibitem [{\citenamefont {Okumura}\ \emph {et~al.}(2019)\citenamefont {Okumura}, \citenamefont {Sato},\ and\ \citenamefont {Kotani}}]{Okumura2019}%
  \BibitemOpen
  \bibfield  {author} {\bibinfo {author} {\bibfnamefont {H.}~\bibnamefont {Okumura}}, \bibinfo {author} {\bibfnamefont {K.}~\bibnamefont {Sato}},\ and\ \bibinfo {author} {\bibfnamefont {T.}~\bibnamefont {Kotani}},\ }\href {https://doi.org/10.1103/PhysRevB.100.054419} {\bibfield  {journal} {\bibinfo  {journal} {Physical Review B}\ }\textbf {\bibinfo {volume} {100}},\ \bibinfo {pages} {054419} (\bibinfo {year} {2019})}\BibitemShut {NoStop}%
\bibitem [{\citenamefont {Friedrich}\ \emph {et~al.}(2020)\citenamefont {Friedrich}, \citenamefont {M{\"{u}}ller},\ and\ \citenamefont {Bl{\"{u}}gel}}]{Friedrich2020}%
  \BibitemOpen
  \bibfield  {author} {\bibinfo {author} {\bibfnamefont {C.}~\bibnamefont {Friedrich}}, \bibinfo {author} {\bibfnamefont {M.~C. T.~D.}\ \bibnamefont {M{\"{u}}ller}},\ and\ \bibinfo {author} {\bibfnamefont {S.}~\bibnamefont {Bl{\"{u}}gel}},\ }in\ \href {https://doi.org/10.1007/978-3-319-44677-6_74} {\emph {\bibinfo {booktitle} {Handbook of Materials Modeling: Methods: Theory and Modeling}}},\ \bibinfo {editor} {edited by\ \bibinfo {editor} {\bibfnamefont {W.}~\bibnamefont {Andreoni}}\ and\ \bibinfo {editor} {\bibfnamefont {S.}~\bibnamefont {Yip}}}\ (\bibinfo  {publisher} {Springer International Publishing},\ \bibinfo {address} {Cham},\ \bibinfo {year} {2020})\ pp.\ \bibinfo {pages} {919--956}\BibitemShut {NoStop}%
\bibitem [{\citenamefont {Skovhus}\ and\ \citenamefont {Olsen}(2021)}]{Skovhus2021}%
  \BibitemOpen
  \bibfield  {author} {\bibinfo {author} {\bibfnamefont {T.}~\bibnamefont {Skovhus}}\ and\ \bibinfo {author} {\bibfnamefont {T.}~\bibnamefont {Olsen}},\ }\href {https://doi.org/10.1103/PhysRevB.103.245110} {\bibfield  {journal} {\bibinfo  {journal} {Physical Review B}\ }\textbf {\bibinfo {volume} {103}},\ \bibinfo {pages} {245110} (\bibinfo {year} {2021})}\BibitemShut {NoStop}%
\bibitem [{\citenamefont {Mook}\ and\ \citenamefont {Paul}(1985)}]{Mook1985}%
  \BibitemOpen
  \bibfield  {author} {\bibinfo {author} {\bibfnamefont {H.~A.}\ \bibnamefont {Mook}}\ and\ \bibinfo {author} {\bibfnamefont {D.~M.}\ \bibnamefont {Paul}},\ }\href {https://doi.org/10.1103/PhysRevLett.54.227} {\bibfield  {journal} {\bibinfo  {journal} {Physical Review Letters}\ }\textbf {\bibinfo {volume} {54}},\ \bibinfo {pages} {227} (\bibinfo {year} {1985})}\BibitemShut {NoStop}%
\bibitem [{\citenamefont {Moriya}(1985)}]{Moriya1985}%
  \BibitemOpen
  \bibfield  {author} {\bibinfo {author} {\bibfnamefont {T.}~\bibnamefont {Moriya}},\ }\href {https://doi.org/10.1007/978-3-642-82499-9} {\emph {\bibinfo {title} {Spin Fluctuations in Itinerant Electron Magnetism}}},\ \bibinfo {series} {Springer Series in Solid-State Sciences}, Vol.~\bibinfo {volume} {56}\ (\bibinfo  {publisher} {Springer-Verlag Berlin Heidelberg},\ \bibinfo {year} {1985})\BibitemShut {NoStop}%
\bibitem [{\citenamefont {Fei}\ \emph {et~al.}(2018)\citenamefont {Fei}, \citenamefont {Huang}, \citenamefont {Malinowski}, \citenamefont {Wang}, \citenamefont {Song}, \citenamefont {Sanchez}, \citenamefont {Yao}, \citenamefont {Xiao}, \citenamefont {Zhu}, \citenamefont {May}, \citenamefont {Wu}, \citenamefont {Cobden}, \citenamefont {Chu},\ and\ \citenamefont {Xu}}]{Fei2018}%
  \BibitemOpen
  \bibfield  {author} {\bibinfo {author} {\bibfnamefont {Z.}~\bibnamefont {Fei}}, \bibinfo {author} {\bibfnamefont {B.}~\bibnamefont {Huang}}, \bibinfo {author} {\bibfnamefont {P.}~\bibnamefont {Malinowski}}, \bibinfo {author} {\bibfnamefont {W.}~\bibnamefont {Wang}}, \bibinfo {author} {\bibfnamefont {T.}~\bibnamefont {Song}}, \bibinfo {author} {\bibfnamefont {J.}~\bibnamefont {Sanchez}}, \bibinfo {author} {\bibfnamefont {W.}~\bibnamefont {Yao}}, \bibinfo {author} {\bibfnamefont {D.}~\bibnamefont {Xiao}}, \bibinfo {author} {\bibfnamefont {X.}~\bibnamefont {Zhu}}, \bibinfo {author} {\bibfnamefont {A.~F.}\ \bibnamefont {May}}, \bibinfo {author} {\bibfnamefont {W.}~\bibnamefont {Wu}}, \bibinfo {author} {\bibfnamefont {D.~H.}\ \bibnamefont {Cobden}}, \bibinfo {author} {\bibfnamefont {J.-H.}\ \bibnamefont {Chu}},\ and\ \bibinfo {author} {\bibfnamefont {X.}~\bibnamefont {Xu}},\ }\href {https://doi.org/10.1038/s41563-018-0149-7} {\bibfield  {journal} {\bibinfo  {journal} {Nature Materials}\ }\textbf {\bibinfo
  {volume} {17}},\ \bibinfo {pages} {778} (\bibinfo {year} {2018})},\ \Eprint {https://arxiv.org/abs/1803.02559} {1803.02559} \BibitemShut {NoStop}%
\bibitem [{\citenamefont {Alghamdi}\ \emph {et~al.}(2019)\citenamefont {Alghamdi}, \citenamefont {Lohmann}, \citenamefont {Li}, \citenamefont {Jothi}, \citenamefont {Shao}, \citenamefont {Aldosary}, \citenamefont {Su}, \citenamefont {Fokwa},\ and\ \citenamefont {Shi}}]{Alghamdi2019}%
  \BibitemOpen
  \bibfield  {author} {\bibinfo {author} {\bibfnamefont {M.}~\bibnamefont {Alghamdi}}, \bibinfo {author} {\bibfnamefont {M.}~\bibnamefont {Lohmann}}, \bibinfo {author} {\bibfnamefont {J.}~\bibnamefont {Li}}, \bibinfo {author} {\bibfnamefont {P.~R.}\ \bibnamefont {Jothi}}, \bibinfo {author} {\bibfnamefont {Q.}~\bibnamefont {Shao}}, \bibinfo {author} {\bibfnamefont {M.}~\bibnamefont {Aldosary}}, \bibinfo {author} {\bibfnamefont {T.}~\bibnamefont {Su}}, \bibinfo {author} {\bibfnamefont {B.~P.}\ \bibnamefont {Fokwa}},\ and\ \bibinfo {author} {\bibfnamefont {J.}~\bibnamefont {Shi}},\ }\href {https://doi.org/10.1021/acs.nanolett.9b01043} {\bibfield  {journal} {\bibinfo  {journal} {Nano Letters}\ }\textbf {\bibinfo {volume} {19}},\ \bibinfo {pages} {4400} (\bibinfo {year} {2019})},\ \Eprint {https://arxiv.org/abs/1903.00571} {1903.00571} \BibitemShut {NoStop}%
\bibitem [{\citenamefont {Wang}\ \emph {et~al.}(2023)\citenamefont {Wang}, \citenamefont {Wu}, \citenamefont {Zhang}, \citenamefont {Liu}, \citenamefont {Chen}, \citenamefont {Pandey}, \citenamefont {Yin}, \citenamefont {Wei}, \citenamefont {Lei}, \citenamefont {Shi}, \citenamefont {Lu}, \citenamefont {Li}, \citenamefont {Fert}, \citenamefont {Wang}, \citenamefont {Nie},\ and\ \citenamefont {Zhao}}]{Wang2023}%
  \BibitemOpen
  \bibfield  {author} {\bibinfo {author} {\bibfnamefont {H.}~\bibnamefont {Wang}}, \bibinfo {author} {\bibfnamefont {H.}~\bibnamefont {Wu}}, \bibinfo {author} {\bibfnamefont {J.}~\bibnamefont {Zhang}}, \bibinfo {author} {\bibfnamefont {Y.}~\bibnamefont {Liu}}, \bibinfo {author} {\bibfnamefont {D.}~\bibnamefont {Chen}}, \bibinfo {author} {\bibfnamefont {C.}~\bibnamefont {Pandey}}, \bibinfo {author} {\bibfnamefont {J.}~\bibnamefont {Yin}}, \bibinfo {author} {\bibfnamefont {D.}~\bibnamefont {Wei}}, \bibinfo {author} {\bibfnamefont {N.}~\bibnamefont {Lei}}, \bibinfo {author} {\bibfnamefont {S.}~\bibnamefont {Shi}}, \bibinfo {author} {\bibfnamefont {H.}~\bibnamefont {Lu}}, \bibinfo {author} {\bibfnamefont {P.}~\bibnamefont {Li}}, \bibinfo {author} {\bibfnamefont {A.}~\bibnamefont {Fert}}, \bibinfo {author} {\bibfnamefont {K.~L.}\ \bibnamefont {Wang}}, \bibinfo {author} {\bibfnamefont {T.}~\bibnamefont {Nie}},\ and\ \bibinfo {author} {\bibfnamefont {W.}~\bibnamefont {Zhao}},\ }\href
  {https://doi.org/10.1038/s41467-023-40714-y} {\bibfield  {journal} {\bibinfo  {journal} {Nature Communications}\ }\textbf {\bibinfo {volume} {14}},\ \bibinfo {pages} {5173} (\bibinfo {year} {2023})}\BibitemShut {NoStop}%
\bibitem [{\citenamefont {Runge}\ and\ \citenamefont {Gross}(1984)}]{Runge1984}%
  \BibitemOpen
  \bibfield  {author} {\bibinfo {author} {\bibfnamefont {E.}~\bibnamefont {Runge}}\ and\ \bibinfo {author} {\bibfnamefont {E.~K.~U.}\ \bibnamefont {Gross}},\ }\href {https://doi.org/10.1103/PhysRevLett.52.997} {\bibfield  {journal} {\bibinfo  {journal} {Physical Review Letters}\ }\textbf {\bibinfo {volume} {52}},\ \bibinfo {pages} {997} (\bibinfo {year} {1984})}\BibitemShut {NoStop}%
\bibitem [{\citenamefont {Gross}\ and\ \citenamefont {Kohn}(1985)}]{Gross1985}%
  \BibitemOpen
  \bibfield  {author} {\bibinfo {author} {\bibfnamefont {E.~K.~U.}\ \bibnamefont {Gross}}\ and\ \bibinfo {author} {\bibfnamefont {W.}~\bibnamefont {Kohn}},\ }\href {https://doi.org/10.1103/PhysRevLett.55.2850} {\bibfield  {journal} {\bibinfo  {journal} {Physical Review Letters}\ }\textbf {\bibinfo {volume} {55}},\ \bibinfo {pages} {2850} (\bibinfo {year} {1985})}\BibitemShut {NoStop}%
\bibitem [{\citenamefont {{Van Hove}}(1954)}]{VanHove1954}%
  \BibitemOpen
  \bibfield  {author} {\bibinfo {author} {\bibfnamefont {L.}~\bibnamefont {{Van Hove}}},\ }\href {https://doi.org/10.1103/PhysRev.95.1374} {\bibfield  {journal} {\bibinfo  {journal} {Physical Review}\ }\textbf {\bibinfo {volume} {95}},\ \bibinfo {pages} {1374} (\bibinfo {year} {1954})}\BibitemShut {NoStop}%
\bibitem [{\citenamefont {Kubo}(1966)}]{Kubo1966}%
  \BibitemOpen
  \bibfield  {author} {\bibinfo {author} {\bibfnamefont {R.}~\bibnamefont {Kubo}},\ }\href {http://iopscience.iop.org/article/10.1088/0034-4885/29/1/306/meta} {\bibfield  {journal} {\bibinfo  {journal} {Rep. Prog. Phys.}\ }\textbf {\bibinfo {volume} {29}},\ \bibinfo {pages} {255} (\bibinfo {year} {1966})}\BibitemShut {NoStop}%
\bibitem [{new()}]{new_implementation}%
  \BibitemOpen
  \href@noop {} {\bibinfo {title} {The details of the implementation will be documented elsewhere.}}\BibitemShut {Stop}%
\bibitem [{\citenamefont {Yan}\ \emph {et~al.}(2011)\citenamefont {Yan}, \citenamefont {Mortensen}, \citenamefont {Jacobsen},\ and\ \citenamefont {Thygesen}}]{Yan2011}%
  \BibitemOpen
  \bibfield  {author} {\bibinfo {author} {\bibfnamefont {J.}~\bibnamefont {Yan}}, \bibinfo {author} {\bibfnamefont {J.~J.}\ \bibnamefont {Mortensen}}, \bibinfo {author} {\bibfnamefont {K.~W.}\ \bibnamefont {Jacobsen}},\ and\ \bibinfo {author} {\bibfnamefont {K.~S.}\ \bibnamefont {Thygesen}},\ }\href {https://doi.org/10.1103/PhysRevB.83.245122} {\bibfield  {journal} {\bibinfo  {journal} {Physical Review B}\ }\textbf {\bibinfo {volume} {83}},\ \bibinfo {pages} {245122} (\bibinfo {year} {2011})}\BibitemShut {NoStop}%
\bibitem [{\citenamefont {Mortensen}\ \emph {et~al.}(2024)\citenamefont {Mortensen}, \citenamefont {Larsen}, \citenamefont {Kuisma}, \citenamefont {Ivanov}, \citenamefont {Taghizadeh}, \citenamefont {Peterson}, \citenamefont {Haldar}, \citenamefont {Dohn}, \citenamefont {Sch{\"a}fer}, \citenamefont {J{\'o}nsson}, \citenamefont {Hermes}, \citenamefont {Nilsson}, \citenamefont {Kastlunger}, \citenamefont {Levi}, \citenamefont {J{\'o}nsson}, \citenamefont {H{\"a}kkinen}, \citenamefont {Fojt}, \citenamefont {Kangsabanik}, \citenamefont {S{\o}dequist}, \citenamefont {Lehtom{\"a}ki}, \citenamefont {Heske}, \citenamefont {Enkovaara}, \citenamefont {Winther}, \citenamefont {Dulak}, \citenamefont {Melander}, \citenamefont {Ovesen}, \citenamefont {Louhivuori}, \citenamefont {Walter}, \citenamefont {Gjerding}, \citenamefont {Lopez-Acevedo}, \citenamefont {Erhart}, \citenamefont {Warmbier}, \citenamefont {W{\"u}rdemann}, \citenamefont {Kaappa}, \citenamefont {Latini}, \citenamefont {Boland}, \citenamefont {Bligaard},
  \citenamefont {Skovhus}, \citenamefont {Susi}, \citenamefont {Maxson}, \citenamefont {Rossi}, \citenamefont {Chen}, \citenamefont {Schmerwitz}, \citenamefont {Schi{\o}tz}, \citenamefont {Olsen}, \citenamefont {Jacobsen},\ and\ \citenamefont {Thygesen}}]{mortensen2024}%
  \BibitemOpen
  \bibfield  {author} {\bibinfo {author} {\bibfnamefont {J.~J.}\ \bibnamefont {Mortensen}}, \bibinfo {author} {\bibfnamefont {A.~H.}\ \bibnamefont {Larsen}}, \bibinfo {author} {\bibfnamefont {M.}~\bibnamefont {Kuisma}}, \bibinfo {author} {\bibfnamefont {A.~V.}\ \bibnamefont {Ivanov}}, \bibinfo {author} {\bibfnamefont {A.}~\bibnamefont {Taghizadeh}}, \bibinfo {author} {\bibfnamefont {A.}~\bibnamefont {Peterson}}, \bibinfo {author} {\bibfnamefont {A.}~\bibnamefont {Haldar}}, \bibinfo {author} {\bibfnamefont {A.~O.}\ \bibnamefont {Dohn}}, \bibinfo {author} {\bibfnamefont {C.}~\bibnamefont {Sch{\"a}fer}}, \bibinfo {author} {\bibfnamefont {E.~{\"O}.}\ \bibnamefont {J{\'o}nsson}}, \bibinfo {author} {\bibfnamefont {E.~D.}\ \bibnamefont {Hermes}}, \bibinfo {author} {\bibfnamefont {F.~A.}\ \bibnamefont {Nilsson}}, \bibinfo {author} {\bibfnamefont {G.}~\bibnamefont {Kastlunger}}, \bibinfo {author} {\bibfnamefont {G.}~\bibnamefont {Levi}}, \bibinfo {author} {\bibfnamefont {H.}~\bibnamefont {J{\'o}nsson}}, \bibinfo
  {author} {\bibfnamefont {H.}~\bibnamefont {H{\"a}kkinen}}, \bibinfo {author} {\bibfnamefont {J.}~\bibnamefont {Fojt}}, \bibinfo {author} {\bibfnamefont {J.}~\bibnamefont {Kangsabanik}}, \bibinfo {author} {\bibfnamefont {J.}~\bibnamefont {S{\o}dequist}}, \bibinfo {author} {\bibfnamefont {J.}~\bibnamefont {Lehtom{\"a}ki}}, \bibinfo {author} {\bibfnamefont {J.}~\bibnamefont {Heske}}, \bibinfo {author} {\bibfnamefont {J.}~\bibnamefont {Enkovaara}}, \bibinfo {author} {\bibfnamefont {K.~T.}\ \bibnamefont {Winther}}, \bibinfo {author} {\bibfnamefont {M.}~\bibnamefont {Dulak}}, \bibinfo {author} {\bibfnamefont {M.~M.}\ \bibnamefont {Melander}}, \bibinfo {author} {\bibfnamefont {M.}~\bibnamefont {Ovesen}}, \bibinfo {author} {\bibfnamefont {M.}~\bibnamefont {Louhivuori}}, \bibinfo {author} {\bibfnamefont {M.}~\bibnamefont {Walter}}, \bibinfo {author} {\bibfnamefont {M.}~\bibnamefont {Gjerding}}, \bibinfo {author} {\bibfnamefont {O.}~\bibnamefont {Lopez-Acevedo}}, \bibinfo {author} {\bibfnamefont {P.}~\bibnamefont
  {Erhart}}, \bibinfo {author} {\bibfnamefont {R.}~\bibnamefont {Warmbier}}, \bibinfo {author} {\bibfnamefont {R.}~\bibnamefont {W{\"u}rdemann}}, \bibinfo {author} {\bibfnamefont {S.}~\bibnamefont {Kaappa}}, \bibinfo {author} {\bibfnamefont {S.}~\bibnamefont {Latini}}, \bibinfo {author} {\bibfnamefont {T.~M.}\ \bibnamefont {Boland}}, \bibinfo {author} {\bibfnamefont {T.}~\bibnamefont {Bligaard}}, \bibinfo {author} {\bibfnamefont {T.}~\bibnamefont {Skovhus}}, \bibinfo {author} {\bibfnamefont {T.}~\bibnamefont {Susi}}, \bibinfo {author} {\bibfnamefont {T.}~\bibnamefont {Maxson}}, \bibinfo {author} {\bibfnamefont {T.}~\bibnamefont {Rossi}}, \bibinfo {author} {\bibfnamefont {X.}~\bibnamefont {Chen}}, \bibinfo {author} {\bibfnamefont {Y.~L.~A.}\ \bibnamefont {Schmerwitz}}, \bibinfo {author} {\bibfnamefont {J.}~\bibnamefont {Schi{\o}tz}}, \bibinfo {author} {\bibfnamefont {T.}~\bibnamefont {Olsen}}, \bibinfo {author} {\bibfnamefont {K.~W.}\ \bibnamefont {Jacobsen}},\ and\ \bibinfo {author} {\bibfnamefont {K.~S.}\
  \bibnamefont {Thygesen}},\ }\href {https://doi.org/10.1063/5.0182685} {\bibfield  {journal} {\bibinfo  {journal} {The Journal of Chemical Physics}\ }\textbf {\bibinfo {volume} {160}},\ \bibinfo {pages} {92503} (\bibinfo {year} {2024})}\BibitemShut {NoStop}%
\bibitem [{\citenamefont {Bl{\"{o}}chl}(1994)}]{Blochl1994}%
  \BibitemOpen
  \bibfield  {author} {\bibinfo {author} {\bibfnamefont {P.~E.}\ \bibnamefont {Bl{\"{o}}chl}},\ }\href {https://doi.org/10.1103/PhysRevB.50.17953} {\bibfield  {journal} {\bibinfo  {journal} {Physical Review B}\ }\textbf {\bibinfo {volume} {50}},\ \bibinfo {pages} {17953} (\bibinfo {year} {1994})}\BibitemShut {NoStop}%
\bibitem [{\citenamefont {Skovhus}(2021)}]{SkovhusPhD}%
  \BibitemOpen
  \bibfield  {author} {\bibinfo {author} {\bibfnamefont {T.}~\bibnamefont {Skovhus}},\ }\emph {\bibinfo {title} {{Magnetic excitations from first principles}}},\ \href@noop {} {Ph.D. thesis},\ \bibinfo  {school} {Department of Physics, Technical University of Denmark} (\bibinfo {year} {2021})\BibitemShut {NoStop}%
\bibitem [{\citenamefont {Costa}\ \emph {et~al.}(2020)\citenamefont {Costa}, \citenamefont {Peres}, \citenamefont {Fern{\'{a}}ndez-Rossier},\ and\ \citenamefont {Costa}}]{Costa2020a}%
  \BibitemOpen
  \bibfield  {author} {\bibinfo {author} {\bibfnamefont {M.}~\bibnamefont {Costa}}, \bibinfo {author} {\bibfnamefont {N.~M.~R.}\ \bibnamefont {Peres}}, \bibinfo {author} {\bibfnamefont {J.}~\bibnamefont {Fern{\'{a}}ndez-Rossier}},\ and\ \bibinfo {author} {\bibfnamefont {A.~T.}\ \bibnamefont {Costa}},\ }\href {https://doi.org/10.1103/PhysRevB.102.014450} {\bibfield  {journal} {\bibinfo  {journal} {Physical Review B}\ }\textbf {\bibinfo {volume} {102}},\ \bibinfo {pages} {014450} (\bibinfo {year} {2020})}\BibitemShut {NoStop}%
\bibitem [{\citenamefont {Liechtenstein}\ \emph {et~al.}(1987)\citenamefont {Liechtenstein}, \citenamefont {Katsnelson}, \citenamefont {Antropov},\ and\ \citenamefont {Gubanov}}]{Liechtenstein1987}%
  \BibitemOpen
  \bibfield  {author} {\bibinfo {author} {\bibfnamefont {A.}~\bibnamefont {Liechtenstein}}, \bibinfo {author} {\bibfnamefont {M.}~\bibnamefont {Katsnelson}}, \bibinfo {author} {\bibfnamefont {V.}~\bibnamefont {Antropov}},\ and\ \bibinfo {author} {\bibfnamefont {V.}~\bibnamefont {Gubanov}},\ }\href {https://doi.org/10.1016/0304-8853(87)90721-9} {\bibfield  {journal} {\bibinfo  {journal} {J. Magn. Magn. Mater.}\ }\textbf {\bibinfo {volume} {67}},\ \bibinfo {pages} {65} (\bibinfo {year} {1987})}\BibitemShut {NoStop}%
\bibitem [{\citenamefont {Bruno}(2003)}]{Bruno2003}%
  \BibitemOpen
  \bibfield  {author} {\bibinfo {author} {\bibfnamefont {P.}~\bibnamefont {Bruno}},\ }\href {https://doi.org/10.1103/PhysRevLett.90.087205} {\bibfield  {journal} {\bibinfo  {journal} {Phys. Rev. Lett.}\ }\textbf {\bibinfo {volume} {90}},\ \bibinfo {pages} {087205} (\bibinfo {year} {2003})}\BibitemShut {NoStop}%
\bibitem [{\citenamefont {Durhuus}\ \emph {et~al.}(2023)\citenamefont {Durhuus}, \citenamefont {Skovhus},\ and\ \citenamefont {Olsen}}]{Durhuus2023}%
  \BibitemOpen
  \bibfield  {author} {\bibinfo {author} {\bibfnamefont {F.~L.}\ \bibnamefont {Durhuus}}, \bibinfo {author} {\bibfnamefont {T.}~\bibnamefont {Skovhus}},\ and\ \bibinfo {author} {\bibfnamefont {T.}~\bibnamefont {Olsen}},\ }\href {https://doi.org/10.1088/1361-648X/acab4b} {\bibfield  {journal} {\bibinfo  {journal} {J. Phys. Condens. Matter}\ }\textbf {\bibinfo {volume} {35}},\ \bibinfo {pages} {105802} (\bibinfo {year} {2023})}\BibitemShut {NoStop}%
\bibitem [{\citenamefont {Ghosh}\ \emph {et~al.}(2023)\citenamefont {Ghosh}, \citenamefont {Ershadrad}, \citenamefont {Borisov},\ and\ \citenamefont {Sanyal}}]{Ghosh2023}%
  \BibitemOpen
  \bibfield  {author} {\bibinfo {author} {\bibfnamefont {S.}~\bibnamefont {Ghosh}}, \bibinfo {author} {\bibfnamefont {S.}~\bibnamefont {Ershadrad}}, \bibinfo {author} {\bibfnamefont {V.}~\bibnamefont {Borisov}},\ and\ \bibinfo {author} {\bibfnamefont {B.}~\bibnamefont {Sanyal}},\ }\href {https://doi.org/10.1038/s41524-023-01024-5} {\bibfield  {journal} {\bibinfo  {journal} {npj Computational Materials}\ }\textbf {\bibinfo {volume} {9}},\ \bibinfo {pages} {86} (\bibinfo {year} {2023})}\BibitemShut {NoStop}%
\bibitem [{\citenamefont {Zakeri}(2014)}]{Zakeri2014}%
  \BibitemOpen
  \bibfield  {author} {\bibinfo {author} {\bibfnamefont {K.}~\bibnamefont {Zakeri}},\ }\href {https://doi.org/10.1016/j.physrep.2014.08.001} {\bibfield  {journal} {\bibinfo  {journal} {Physics Reports}\ }\textbf {\bibinfo {volume} {545}},\ \bibinfo {pages} {47} (\bibinfo {year} {2014})}\BibitemShut {NoStop}%
\bibitem [{\citenamefont {Qin}\ \emph {et~al.}(2015)\citenamefont {Qin}, \citenamefont {Zakeri}, \citenamefont {Ernst}, \citenamefont {Sandratskii}, \citenamefont {Buczek}, \citenamefont {Marmodoro}, \citenamefont {Chuang}, \citenamefont {Zhang},\ and\ \citenamefont {Kirschner}}]{Qin2015}%
  \BibitemOpen
  \bibfield  {author} {\bibinfo {author} {\bibfnamefont {H.~J.}\ \bibnamefont {Qin}}, \bibinfo {author} {\bibfnamefont {K.}~\bibnamefont {Zakeri}}, \bibinfo {author} {\bibfnamefont {A.}~\bibnamefont {Ernst}}, \bibinfo {author} {\bibfnamefont {L.~M.}\ \bibnamefont {Sandratskii}}, \bibinfo {author} {\bibfnamefont {P.}~\bibnamefont {Buczek}}, \bibinfo {author} {\bibfnamefont {A.}~\bibnamefont {Marmodoro}}, \bibinfo {author} {\bibfnamefont {T.~H.}\ \bibnamefont {Chuang}}, \bibinfo {author} {\bibfnamefont {Y.}~\bibnamefont {Zhang}},\ and\ \bibinfo {author} {\bibfnamefont {J.}~\bibnamefont {Kirschner}},\ }\href {https://doi.org/10.1038/ncomms7126} {\bibfield  {journal} {\bibinfo  {journal} {Nature Communications}\ }\textbf {\bibinfo {volume} {6}},\ \bibinfo {pages} {6126} (\bibinfo {year} {2015})}\BibitemShut {NoStop}%
\bibitem [{\citenamefont {Wang}\ and\ \citenamefont {Zhang}(2023)}]{Wang2023a}%
  \BibitemOpen
  \bibfield  {author} {\bibinfo {author} {\bibfnamefont {F.}~\bibnamefont {Wang}}\ and\ \bibinfo {author} {\bibfnamefont {H.}~\bibnamefont {Zhang}},\ }\href {https://doi.org/10.1103/PhysRevB.108.195140} {\bibfield  {journal} {\bibinfo  {journal} {Physical Review B}\ }\textbf {\bibinfo {volume} {108}},\ \bibinfo {pages} {195140} (\bibinfo {year} {2023})}\BibitemShut {NoStop}%
\bibitem [{\citenamefont {Haastrup}\ \emph {et~al.}(2018)\citenamefont {Haastrup}, \citenamefont {Strange}, \citenamefont {Pandey}, \citenamefont {Deilmann}, \citenamefont {Schmidt}, \citenamefont {Hinsche}, \citenamefont {Gjerding}, \citenamefont {Torelli}, \citenamefont {Larsen}, \citenamefont {Riis-Jensen}, \citenamefont {Gath}, \citenamefont {Jacobsen}, \citenamefont {Mortensen}, \citenamefont {Olsen},\ and\ \citenamefont {Thygesen}}]{Haastrup2018}%
  \BibitemOpen
  \bibfield  {author} {\bibinfo {author} {\bibfnamefont {S.}~\bibnamefont {Haastrup}}, \bibinfo {author} {\bibfnamefont {M.}~\bibnamefont {Strange}}, \bibinfo {author} {\bibfnamefont {M.}~\bibnamefont {Pandey}}, \bibinfo {author} {\bibfnamefont {T.}~\bibnamefont {Deilmann}}, \bibinfo {author} {\bibfnamefont {P.~S.}\ \bibnamefont {Schmidt}}, \bibinfo {author} {\bibfnamefont {N.~F.}\ \bibnamefont {Hinsche}}, \bibinfo {author} {\bibfnamefont {M.~N.}\ \bibnamefont {Gjerding}}, \bibinfo {author} {\bibfnamefont {D.}~\bibnamefont {Torelli}}, \bibinfo {author} {\bibfnamefont {P.~M.}\ \bibnamefont {Larsen}}, \bibinfo {author} {\bibfnamefont {A.~C.}\ \bibnamefont {Riis-Jensen}}, \bibinfo {author} {\bibfnamefont {J.}~\bibnamefont {Gath}}, \bibinfo {author} {\bibfnamefont {K.~W.}\ \bibnamefont {Jacobsen}}, \bibinfo {author} {\bibfnamefont {J.~J.}\ \bibnamefont {Mortensen}}, \bibinfo {author} {\bibfnamefont {T.}~\bibnamefont {Olsen}},\ and\ \bibinfo {author} {\bibfnamefont {K.~S.}\ \bibnamefont {Thygesen}},\ }\href
  {https://doi.org/10.1088/2053-1583/aacfc1} {\bibfield  {journal} {\bibinfo  {journal} {2D Materials}\ }\textbf {\bibinfo {volume} {5}},\ \bibinfo {pages} {042002} (\bibinfo {year} {2018})}\BibitemShut {NoStop}%
\bibitem [{\citenamefont {Gjerding}\ \emph {et~al.}(2021)\citenamefont {Gjerding}, \citenamefont {Taghizadeh}, \citenamefont {Rasmussen}, \citenamefont {Ali}, \citenamefont {Bertoldo}, \citenamefont {Deilmann}, \citenamefont {Kn{\o}sgaard}, \citenamefont {Kruse}, \citenamefont {Larsen}, \citenamefont {Manti}, \citenamefont {Pedersen}, \citenamefont {Petralanda}, \citenamefont {Skovhus}, \citenamefont {Svendsen}, \citenamefont {Mortensen}, \citenamefont {Olsen},\ and\ \citenamefont {Thygesen}}]{Gjerding2021}%
  \BibitemOpen
  \bibfield  {author} {\bibinfo {author} {\bibfnamefont {M.~N.}\ \bibnamefont {Gjerding}}, \bibinfo {author} {\bibfnamefont {A.}~\bibnamefont {Taghizadeh}}, \bibinfo {author} {\bibfnamefont {A.}~\bibnamefont {Rasmussen}}, \bibinfo {author} {\bibfnamefont {S.}~\bibnamefont {Ali}}, \bibinfo {author} {\bibfnamefont {F.}~\bibnamefont {Bertoldo}}, \bibinfo {author} {\bibfnamefont {T.}~\bibnamefont {Deilmann}}, \bibinfo {author} {\bibfnamefont {N.~R.}\ \bibnamefont {Kn{\o}sgaard}}, \bibinfo {author} {\bibfnamefont {M.}~\bibnamefont {Kruse}}, \bibinfo {author} {\bibfnamefont {A.~H.}\ \bibnamefont {Larsen}}, \bibinfo {author} {\bibfnamefont {S.}~\bibnamefont {Manti}}, \bibinfo {author} {\bibfnamefont {T.~G.}\ \bibnamefont {Pedersen}}, \bibinfo {author} {\bibfnamefont {U.}~\bibnamefont {Petralanda}}, \bibinfo {author} {\bibfnamefont {T.}~\bibnamefont {Skovhus}}, \bibinfo {author} {\bibfnamefont {M.~K.}\ \bibnamefont {Svendsen}}, \bibinfo {author} {\bibfnamefont {J.~J.}\ \bibnamefont {Mortensen}}, \bibinfo {author}
  {\bibfnamefont {T.}~\bibnamefont {Olsen}},\ and\ \bibinfo {author} {\bibfnamefont {K.~S.}\ \bibnamefont {Thygesen}},\ }\href {https://doi.org/10.1088/2053-1583/ac1059} {\bibfield  {journal} {\bibinfo  {journal} {2D Materials}\ }\textbf {\bibinfo {volume} {8}},\ \bibinfo {pages} {044002} (\bibinfo {year} {2021})}\BibitemShut {NoStop}%
\bibitem [{\citenamefont {Lounis}\ \emph {et~al.}(2011)\citenamefont {Lounis}, \citenamefont {Costa}, \citenamefont {Muniz},\ and\ \citenamefont {Mills}}]{Lounis2011}%
  \BibitemOpen
  \bibfield  {author} {\bibinfo {author} {\bibfnamefont {S.}~\bibnamefont {Lounis}}, \bibinfo {author} {\bibfnamefont {A.~T.}\ \bibnamefont {Costa}}, \bibinfo {author} {\bibfnamefont {R.~B.}\ \bibnamefont {Muniz}},\ and\ \bibinfo {author} {\bibfnamefont {D.~L.}\ \bibnamefont {Mills}},\ }\href {https://doi.org/10.1103/PhysRevB.83.035109} {\bibfield  {journal} {\bibinfo  {journal} {Physical Review B}\ }\textbf {\bibinfo {volume} {83}},\ \bibinfo {pages} {035109} (\bibinfo {year} {2011})}\BibitemShut {NoStop}%
\end{thebibliography}%

\end{document}